\documentclass[aps,pra,reprint,superscriptaddress]{revtex4-1}

\usepackage{amsmath}
\usepackage{cases}
\usepackage{graphicx}
\usepackage{mathtools} 
\usepackage[T1]{fontenc} 
\usepackage[utf8]{inputenc} 
\usepackage{amsfonts} 
\usepackage{bm}

\usepackage{color}  
\usepackage{enumitem}
\usepackage{multirow}
\usepackage{tabularx}
\usepackage{longtable}
\usepackage{float}

\usepackage{todonotes} 
\usepackage{xcolor} 

\usepackage{qcircuit}

\newcommand{\QuArC}{%
    \affiliation{%
        Quantum Architectures and Computation Group,
        Microsoft Research,
        Redmond WA 98052, USA.}}
\newcommand{\PNNL}{%
    \affiliation{%
        William R. Wiley Environmental Molecular Sciences Laboratory, Battelle, 
        Pacific Northwest National Laboratory, K8-91, P.O. Box 999, Richland WA 99352, USA}}

\begin{document}


\title{Downfolding of many-body Hamiltonians using active-space models: 
extension of the sub-system embedding sub-algebras approach to unitary  coupled cluster 
formalisms}

\author{Nicholas P. Bauman} \PNNL
\author{Eric J. Bylaska} \PNNL
\author{Sriram Krishnamoorthy} \PNNL
\author{Guang Hao Low} \QuArC
\author{Nathan Wiebe} \QuArC
\author{Karol Kowalski} 
\email{karol.kowalski@pnnl.gov}
\PNNL


\date{\today}

\begin{abstract}
In this paper we outline the extension of recently introduced the sub-system embedding 
sub-algebras coupled cluster (SES-CC) formalism to the unitary CC formalism. 
In analogy to the standard single-reference SES-CC formalism, its 
unitary CC extension allows one to include the dynamical (outside the active space) 
correlation effects in an SES induced complete active space (CAS) effective Hamiltonian. 
In contrast to the standard single-reference SES-CC theory, the unitary CC approach 
results in a Hermitian form of the effective Hamiltonian. Additionally, for the double unitary CC formalism (DUCC)
the corresponding CAS 
eigenvalue problem provides a rigorous separation of external cluster amplitudes that 
describe dynamical correlation effects -- used to define the effective Hamiltonian -- 
from those corresponding to the internal (inside the active space) excitations that 
define the components of eigenvectors associated with the energy of the entire system. 
The proposed formalism can be viewed as an efficient way of downfolding many-electron 
Hamiltonian to the low-energy model represented by a particular choice of CAS. In 
principle, this technique can be extended to any type of complete active space 
representing an arbitrary energy window of a quantum system. The Hermitian character of low-dimensional effective Hamiltonians makes them an ideal target for several types of full configuration interaction 
(FCI) type eigensolvers.
As an example,  we also discuss the algebraic form of the 
perturbative expansions of the effective DUCC Hamiltonians corresponding to composite 
unitary CC theories and discuss possible algorithms for hybrid classical and quantum computing.
\end{abstract}


\maketitle

\section{Introduction}


Even though quantum chemistry and materials science 
communities have expended a great deal of effort designing numerous methods to describe 
collective behavior of electrons in correlated systems, the fundamental understanding 
of these processes in many systems is still inhibited by an exponential growth in computation  
associated with representing the many-body electronic wave function. 
These problems  
usually occur for  systems characterized by small energy gaps between occupied 
and unoccupied one-particle states, where the use of advanced tools to describe 
electron correlation effects is a prerequisite. Several classes of many-body approaches based on the 
inclusion of higher-rank collective excitations~\cite{paldus07,bartlett07_291}, multi-reference
concepts~\cite{doi:10.1080/00268977500103351,LINDGREN198793,jezmonk,ims1,
kaldor1991fock,mahapatra1,meissner1998fock,connectivity1}, and symmetry-breaking 
mechanisms have resulted in a myriad of approximations trying to describe static and 
dynamic correlations effects.
For example, coupled cluster 
(CC)~\cite{coester58_421,coester60_477,cizek66_4256,shavitt72,purvis82_1910},
density matrix renormalization group (DMRG)~\cite{white1992density,
schollwock2005density, chan2011density}, and density matrix methods~\cite{mazziotti2001uncertainty,
mazziotti2012structure} have already demonstrated their efficiency in coping with 
complicated electron correlation effects in modest size molecular and materials systems. However, 
the applicability of these methods for larger systems is still defined by a trade-off between accuracy and computational costs. In many cases, this situation evolves into a scientific stalemate. 
While strongly correlated systems successfully elude the mainstream theoretical 
modeling exemplified by low-rank methods, mainly DFT, the applicability of very 
accurate yet very expensive many-body formalisms is significantly limited by 
computational resources offered by conventional computers. In this regard, 
enabling  mathematically rigorous models where correlation effects are downfolded into a low-dimensionality space 
offers  a unique chance to permanently eliminate the inherent bias/biases of 
currently employed many-body theories.  The diagonalization of the resulting low-dimensionality 
effective Hamiltonians is also an ideal target for various algorithms  including algorithms for
classical computers as well as novel  algorithms capable of taking advantage of emerging quantum
information systems.~\cite{bravyi2000fermionic,
seeley2012bravyi,poulin2017fast,setia2018bravyi,wecker2015progress,low_depth_Chan,motta2018low,mcclean2016theory,
PhysRevA.95.020501}

\begin{figure}
	\includegraphics[angle=0, width=0.49\textwidth]{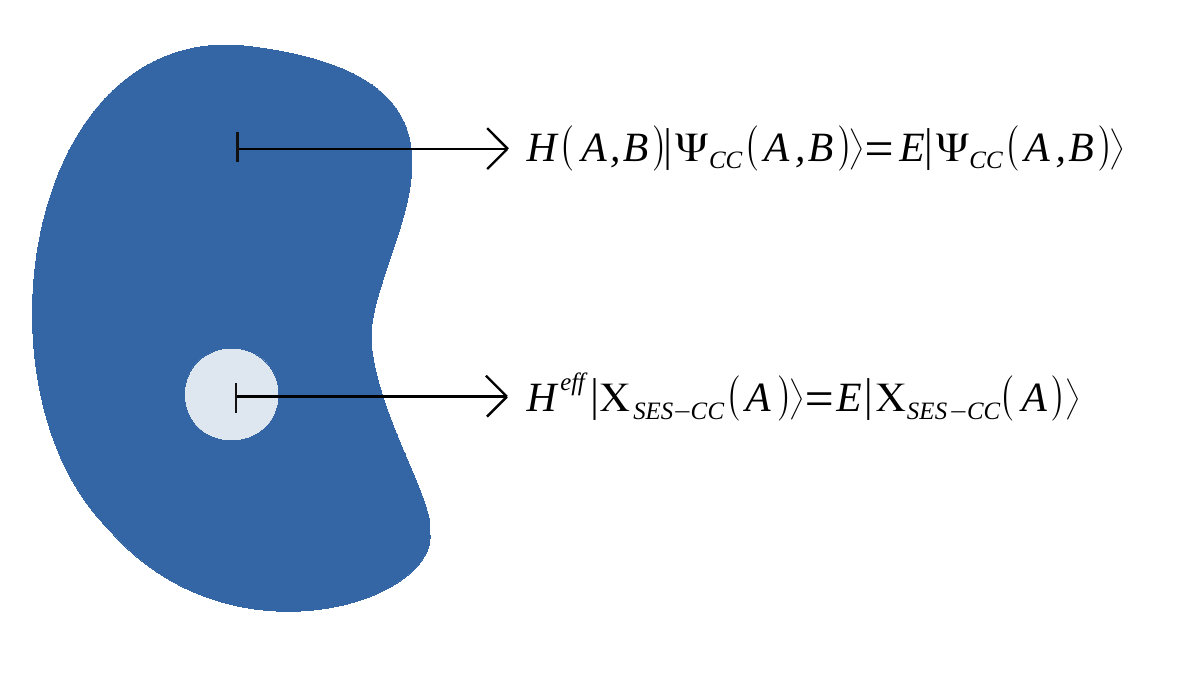}
	\caption{Abstract representation of the SES-CC approach.}
\label{sesabstract}
\end{figure}

Of the various quantum chemistry methods, CC theory has
become the {\it de facto} standard high accuracy calculations for nuclear, atomic, and 
molecular systems. The framework of CC theory, when combined with recent developments of our own,
makes it well suited to address systems in which correlation effects are downfolded onto smaller spaces.
We have recently shown in Refs.~\cite{safkk,kowalski2018regularized} that we can go beyond  solving the ground-state CC equations in the conventional iterative manner, by decoupling the excitations into two disjoint sets as shown in Fig. 1.  The $A$ set is obtained from  predefined classes of excitations (or sub-algebra(s)), and the $B$ set contains all the remaining parameters needed to describe the whole system.  This technique,
known as sub-system embedding sub-algebras coupled cluster (SES-CC) introduces the concept
of active spaces in a natural way and provides a mathematically rigorous procedure for downfolding 
many-body effects for a subset of excitations into an effective Hamiltonian for tractable
eigenvalue problems that provide the ground state energy for the full conventional CC calculation 
(see Fig. \ref{sesabstract}).
The authors of this paper view the SES formalism, which has been shown to work well
compared to standard CC approximations, as potential generalized extension of previously developed embedding methods (see Ref.~\cite{inglesfield1981}) and L\"owdin partitioning techniques~\cite{lowdin1,lowdin2}  which will be discussed in future work. However, as a consequence of the single-reference CC Ansatz, 
the standard SES-CC effective Hamiltonians are not Hermitian, which precludes them from efficient 
utilization with full configuration interaction (FCI) type  diagonalizers, such as those used in 
quantum algorithms and the DMRG method. So there is a natural need for extending SES-CC approach 
to the unitary CC formalisms which assure the 
Hermitian character of resulting effective Hamiltonians. 
Several key problems associated with this extension are related to the  following questions: 
 (1) what is the optimal way of utilizing  diagonalizers 
for enabling  accurate many-body formulations given current and near-term time perspectives? (2) How 
can many-body formulations be tuned (using mathematically rigorous procedures and operating 
on the parameters defining a given many-body approach) to existing and future computational systems 
to assure applications to more realistic and challenging problems than currently possible?

We address the above questions by developing a new 
downfolding strategy pertaining to the SES-CC formalism which enables one to represent the 
effective (or downfolded) many-body Hamiltonian in a much smaller orbital space (or 
active space), where the correlation effects from outside the active space are included in the 
form of a similarity unitary transformation involving only parameters corresponding to high-energy 
components of the corresponding wave function.
The proposed computational scheme can give rise to an efficient  
hybrid classical/quantum computational approach:  while the high-energy components of the wave function 
and the second quantized form of the effective  Hamiltonian are calculated using 
classical computers, 
the diagonalization of the effective  Hamiltonian is achieved by 
employing quantum algorithms. As a specific example of the above formalism, in this paper 
we will consider a formalism where similarity unitary transformation is defined to  
 downfolding an essential part of the virtual orbital 
space. We will also  introduce models based on the inclusion of single and double excitations in 
relevant unitary operators and derive the algebraic form of downfolded Hamiltonian by combining 
particle-hole and physical vacuum representations of second quantized operators.


\section{Standard single-reference formulation}

The standard single-reference CC formulations are predicated on the assumption that 
there exists a reasonable choice of a Slater determinant $|\Phi\rangle$ that can provide 
an adequate starting point for the exponential (CC) expansion of the ground-state 
electronic wave function $|\Psi\rangle$
\begin{equation}
	|\Psi\rangle = e^T|\Phi\rangle \;,
\label{ccp}
\end{equation}
where the so-called cluster operator $T$ can be expanded in terms of its many-body 
components $T_k$
\begin{equation}
	T=\sum_{k=1}^m T_k \;.
\label{tcomp}
\end{equation}
In the second quantized form, each $T_k$ component can be expressed as 
\begin{equation}
	T_k = \frac{1}{(k!)^2} \sum_{\substack{i_1,\ldots,i_k ; \\ a_1, \ldots, a_k}} 
    t_{i_1\ldots i_k}^{a_1\ldots a_k} E^{a_1\ldots a_k}_{i_1\ldots i_k} ~,
\label{xex}
\end{equation}
where indices $i_1,i_2,\ldots$ ($a_1,a_2,\ldots$) refer to occupied (unoccupied) 
spin orbitals in the reference function $|\Phi\rangle$, $t^{i_1\ldots i_k}_{a_1\ldots a_k}$ 
are cluster amplitudes, and $E^{a_1\ldots a_k}_{i_1\ldots i_k} $ are excitation 
operators. The excitation operators are defined through strings of standard creation 
($a_p^\dagger$) and annihilation operators ($a_p$)
\begin{equation}
	E^{a_1\ldots a_k}_{i_1\ldots i_k} = a_{a_1}^{\dagger}\ldots a_{a_k}^{\dagger} 
    a_{i_k}\ldots a_{i_1} \;.
\label{estring}
\end{equation}
In the particle-hole formalism, the excitation operators are expressed in terms of particle/hole creation operators only. 
When acting on the reference function $|\Phi\rangle$, the $E^{a_1\ldots a_k}_{i_1\ldots 
i_k}$ operators produce the so-called excited configurations $|\Phi_{i_1\ldots 
i_k}^{a_1\ldots a_k} \rangle$ defined as
\begin{equation}
	|\Phi_{i_1\ldots i_k}^{a_1\ldots a_k} \rangle = E^{a_1\ldots a_k}_{i_1\ldots i_k} 
    |\Phi\rangle \;.
\label{exconf}
\end{equation}
Upon the substitution of the ansatz (\ref{ccp}) into the Schr\"odinger equation 
one gets the energy-dependent form of the CC equations:
\begin{equation}
	(P+Q) He^T|\Phi\rangle= E (P+Q) e^T |\Phi\rangle \;\;,
\label{schreq}
\end{equation}
where $P$ and $Q$ are projection operators onto the reference function ($P=|\Phi\rangle
\langle\Phi|$) and onto the excited configurations generated by the $T$ operator when 
acting on the reference function, 
\begin{equation}
	Q=\sum_{k=1}^{m}\;\sum_{\substack{i_1<\ldots<i_k; \\ a_1<\ldots <a_k}}
    |\Phi_{i_1\ldots i_k}^{a_1 \ldots a_k}\rangle\langle \Phi_{i_1\ldots i_k}^{a_1 
    \ldots a_k}| \;.
\label{qoper}
\end{equation}
A careful diagrammatic analysis~\cite{paldus07} leads to an equivalent ({\it at the 
solution}), energy-independent form of the CC equations for the cluster amplitudes 
\begin{equation}
	Qe^{-T}He^T|\Phi\rangle = Q\bar{H}|\Phi\rangle = 0 \;,
\label{conform}
\end{equation}
where the $\bar{H}=e^{-T}He^{T}$ is referred to as the similarity transformed Hamiltonian. It can also be shown that the $\bar{H}$ is expressible in terms of connected diagrams, i.e., $\bar{H}=(He^T)_C$,
where $C$ designates a connected form of a given operator expression. Once the $T$ 
operator is determined by solving Eq. (\ref{conform}), the corresponding energy is given 
by the standard formula
\begin{equation}
	E=\langle\Phi| \bar{H}|\Phi\rangle \;.
\label{ccene}
\end{equation}
The second-quantized form of the many-body Hamiltonian defined by up to pairwise 
interactions is given by the formula 
\begin{equation}
	H=\sum_{p,q} h^p_q a_p^{\dagger} a_q + \frac{1}{4}\sum_{p,q,r,s} v^{pq}_{rs} 
    a_p^{\dagger} a_q^{\dagger} a_s a_r \;,
\label{ham}
\end{equation}
where 
$p$,  $q$, $r$, $s$ indices run over a spin-orbital basis involved in a given algebraic 
approximation (associated with the use of finite-dimensional one-particle basis set to 
discretize the Schr\"odinger equation), and $h^p_q$ and $v^{pq}_{rs}$ represent one- and two-electron integrals 
(in the above representation, the  $v^{pq}_{rs}$ tensor is antisymmetric with respect to permutations among the sets of lower and upper spin-orbital indices).  
In typical molecular applications,  based for example
on the delocalized Hartree-Fock molecular orbitals expanded in the Gaussian basis set, the 
number of terms defining second-quantized Hamiltonian is proportional to $N^4$ where 
$N$ stands for the number of basis set functions. It has recently been shown that 
using a different kind of basis set, namely the plane wave dual basis set, the number 
of terms can be reduced from $N^4$ to $N^2$~\cite{low_depth_Chan}. 
Similar reduction can also be achieved for Gaussian basis sets when combined Cholesky and singular value decompositions are employed to represent two-electron integrals (see Refs.~\cite{peng2017highly,motta2018low} for details)

From the point of quantum computing 
applications, the net effect of the number of basis set functions and the number of 
non-vanishing terms in Hamiltonian define the circuit depth that determines the 
efficiency of quantum algorithms. The reduction in the number of non-negligible terms 
may also be achieved by employing localization techniques for Gaussian basis 
sets~\cite{Foster_Boys,Edmiston63,Pipek_Mezey}. An interesting alternative to the 
localized basis sets, that may especially impact the choice of the initial state, is 
the use of the Br\"uckner orbitals~\cite{Brueckner1956,Nesbet1958,Stolarczyk267,
Kobayashi1994,Nooijen_BO,Crawford_BO} that maximize the overlap $|\langle\Phi_B|\Psi\rangle|$ between normalized 
lowest energy Slater determinant $|\Phi_B\rangle$ and the correlated wave function 
$|\Psi\rangle$. Given the form of this condition, one should also expect more efficient 
utilization of phase estimation techniques when Br\"uckner orbitals are employed in the context of various quantum algorithms such as Trotterization. 


In the exact wave function limit, the excitation level $m$ is equal to the number of 
correlated electrons ($N_e$) while in the approximate CC formulations $m\ll N_e$. In this 
way, one can define standard approximations 
such as CCSD ($m=2$)~\cite{purvis82_1910}, CCSDT ($m=3$)~\cite{ccsdt_noga,
ccsdt_noga_err,scuseria_ccsdt}, CCSDTQ ($m=4$)~\cite{Kucharski1991,ccsdtq_nevin}, etc. 
Various standard CC approximations have been successfully applied to describe various 
many-body systems across energy and spatial scales ranging from nuclear matter to molecular 
and extended/periodic systems~\cite{PhysRevC.69.054320,PhysRevLett.92.132501,
PhysRevLett.101.092502,PhysRevLett.117.172501,hirata2001highly,hirata2004coupled,
booth2013towards,mcclain2017gaussian}. The success of the CC methods in capturing 
correlation effects can be attributed to two main factors: (1) its size-extensivity, i.e., 
proper scaling of the energy with number of the particles, which is a direct consequence of 
connected character of diagrams contributing to the CC equations and (2) possibility 
of approximating higher-order excitations by products of low-rank cluster operators.

\section{Properties of CC sub-system embedding sub-algebras}
\label{sessec}
Certain properties of CC equations are inextricably associated with the possibility of 
partitioning of cluster operators in CC wave function into components corresponding to 
various sub-algebras of 
excitation Lie algebra denoted here as
$\mathfrak{g}^{(N)}$, which is generated by all excitation operators 
$E^{a_l}_{i_l}=a_{a_l}^{\dagger} a_{i_l}$. In a recent paper~\cite{safkk}, we have 
analyzed properties of the CC equations stemming from the presence of CC-approximation-specific sub-algebras of excitations that can naturally be identified with the 
so-called active spaces that are frequently used in many areas of quantum chemistry and 
physics. 
Algebraic properties of these sub-algebras  provide a means to re-cast the CC equations in the form of a set 
of eigenvalue problems and a set of equations that couple these eigenvalue problems. Although 
there were several attempts to re-express the CC equations as a non-linear eigenvalue 
problem (either in the context of dressed configuration interaction 
Hamiltonian~\cite{Sanchez-Marin1996}, inclusion of high-order 
excitations~\cite{kallay2000}, or the 
analysis of multiple solutions of CC equations~\cite{zivkovic}), in contrast to 
earlier efforts, all parameters (a subset of the cluster amplitudes) defining the 
matrices to be diagonalized in the eigenvalue subproblems are entirely decoupled from 
those parameters (also a subset of the cluster amplitudes) that define components of 
the corresponding eigenvectors. In particular, it was shown that through the similarity 
transformation of the electronic Hamiltonian it is possible to downfold it to the form 
that acts in the active space and provides corresponding CC energy as its eigenvalue 
value. In contrast to the full electronic Hamiltonian, its effective active-space normal-product form 
representation involves only creation/annihilation operators carrying actives-space 
indices. In typical applications the number of active 
spin orbitals ($N_{\rm act}$) is significantly smaller compared to the total number of 
spin orbitals $N_{\rm S}$, i.e., $N_{\rm act} \ll N_{\rm S}$. 

\begin{figure}
	\includegraphics[angle=0, width=0.50\textwidth]{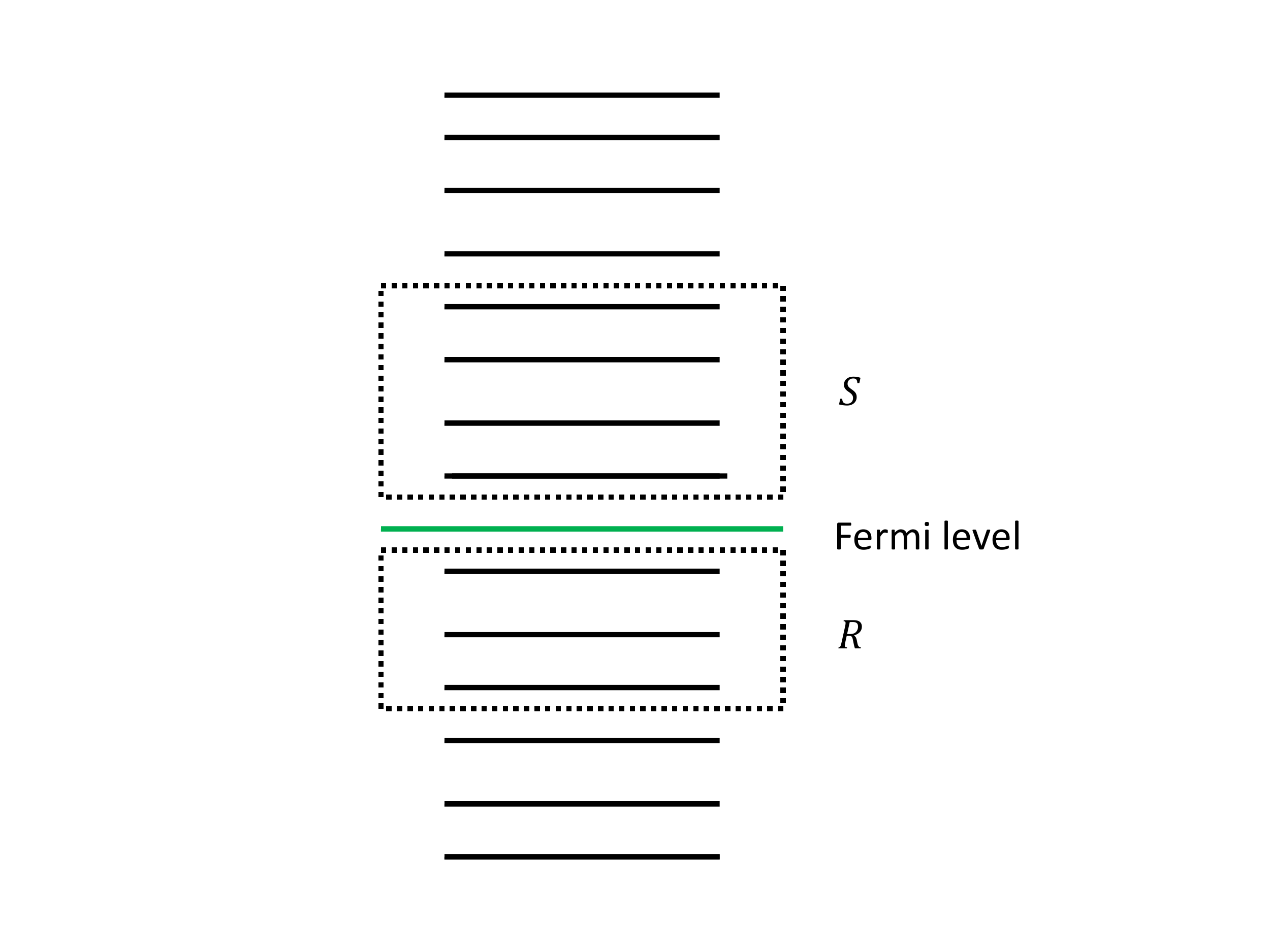}
	\caption{An example of the  $\mathfrak{g}^{(N)}(3_{\rm R},4_{\rm S})$ sub-algebra. As shown in Ref.~\cite{safkk} this sub-algebra is a SES for the CC approach with singles, doubles, triples, quadruples, pentuples, and hextuples (CCSDTQPH).}
\label{fig-1}
\end{figure}

Let us start the discussion by introducing basic notions defining sub-algebras. An 
important class of sub-algebras of the $\mathfrak{g}^{(N)}$ excitation algebra is 
closely related to ideas underlying the active-space concepts in quantum chemistry, 
where one can define sub-algebras $\mathfrak{h}$ through all possible excitations 
$E^{a_1\ldots a_m}_{i_1\ldots i_m}$ that excite electrons from a subset of active 
occupied orbitals (denoted as R) to a subset of active virtual orbitals (denoted as S). 
These sub-algebras will be denoted as $\mathfrak{g}^{(N)}({\rm R,S})$. The 
$\mathfrak{g}^{(N)}({\rm R,S})$ sub-algebras can also be viewed as generators of 
various complete active spaces (CAS(R,S)) spanned by the reference function $|\Phi\rangle$ 
and all excited configurations obtained by acting with elements of 
$\mathfrak{g}^{(N)}({\rm R,S})$ onto $|\Phi\rangle$ (see Fig. \ref{fig-1}). Specific 
examples of these sub-algebras contain sub-algebras involving all occupied and selected 
(S) virtual orbitals or selected set of occupied orbitals (R) and all virtual orbitals, 
which will be denoted for short as $\mathfrak{g}^{(N)}({\rm S})$ and 
$\mathfrak{g}^{(N)}({\rm R})$, respectively. In an alternative notation, we will denote 
$\mathfrak{g}^{(N)}({\rm R},{\rm S})$ as $\mathfrak{g}^{(N)}(x_{\rm R},y_{\rm S})$ 
where numbers $x$ and $y$ refer to the number of orbitals included in sets R and S, 
respectively. Special sub-algebras $\mathfrak{g}^{(N)}({\rm R})$ and 
$\mathfrak{g}^{(N)}({\rm S})$ will be denoted as $\mathfrak{g}^{(N)}(x_{\rm R})$ and 
$\mathfrak{g}^{(N)}(y_{\rm S})$. In this paper, we will entirely focus on active-space 
excitations sub-algebras for the closed-shell single-reference CC formulations. 

An important property of the  excitation sub-algebras is the fact, that in 
general, an arbitrary cluster operator $T$ can be decomposed in a part that belongs to 
sub-algebra of interest $\mathfrak{h}$  (denoted as an internal part, 
$T_{\rm int}(\mathfrak{h})$) and a part that belongs to $\mathfrak{g}^{(N)}-
\mathfrak{h}$ (denoted as an external part, $T_{\rm ext}(\mathfrak{h})$), {\it i.e.}, 
\begin{equation}
	T=T_{\rm int}(\mathfrak{h})+T_{\rm ext}(\mathfrak{h}) \;.
\label{decni}
\end{equation}
The $T_{\rm int}$ operator is a generator of the CAS(R,S) and the corresponding 
amplitudes of the $T_{k,{\rm int}}$ components are labeled exclusively by
active-space spin-orbital indices, while the indices for the  $T_{k,{\rm ext}}$
components of $T_{\rm ext}$ consist of one or more spin-orbitals outside of the active
space. This decomposition entails decomposition of the corresponding CC wave function:
\begin{eqnarray}
	|\Psi\rangle &=& e^{T}|\Phi\rangle = e^{T_{\rm ext}(\mathfrak{h})
    + T_{\rm int}(\mathfrak{h})}|\Phi\rangle  \nonumber \\
	&=& e^{T_{\rm ext}(\mathfrak{h})} e^{T_{\rm int}(\mathfrak{h})}|\Phi\rangle
	= e^{T_{\rm ext}(\mathfrak{h})} |\Psi(\mathfrak{h})\rangle \;,
\label{nildec_1}
\end{eqnarray}
where the CAS-type wave function $|\Psi(\mathfrak{h})\rangle$ is defined as 
\begin{equation}
	|\Psi(\mathfrak{h})\rangle=e^{T_{\rm int}(\mathfrak{h})}|\Phi\rangle \;.
\label{subwave}
\end{equation}
In~\eqref{nildec_1}, we use the fact that $[T_{\rm ext},T_{\rm int}]=0$, which can be seen immediately from the fact that the terms in the cluster operator in particle hole representation can be expressed as $E_{i_1,\ldots,i_k}^{a_1,\ldots,a_k}= b_{a_1}^\dagger \cdots b_{a_k}^\dagger b^{\dagger}_{i_k} \cdots b^{\dagger}_{i_1}$, for anti-commuting $b_\ell^\dagger$, 
where $b_\ell$/$b_\ell^{\dagger}$ operators are defined as
\begin{equation}
b_\ell = 
\begin{cases}
    a_\ell                   &  \ell\in V\\
    a_\ell^{\dagger}   & \ell\in O
  \end{cases}
\;\;,\;\;
b_\ell^{\dagger} = 
\begin{cases}
    a_\ell                   &  \ell\in O\\
    a_\ell^{\dagger}   & \ell\in V
\end{cases}
\;,
\label{bpo}
\end{equation}
and $O$ and $V$ represent sets of occupied and unoccupied spin orbitals in $|\Phi\rangle$, respectively.

The sub-algebras $\mathfrak{h}$ that satisfy two important requirements:
\begin{enumerate}
	\item The $|\Psi(\mathfrak{h})\rangle= e^{T_{\rm int}(\mathfrak{h})}|\Phi\rangle$ is 
    characterized by the same symmetry properties as $|\Psi\rangle$ and $|\Phi\rangle$ 	
    vectors (for example, spin and spatial symmetries). 
	\item The $e^{T_{\rm int}(\mathfrak{h})}|\Phi\rangle$ Ansatz generates the FCI 
    expansion for the subsystem defined by the CAS corresponding to the $\mathfrak{h}$ 
    sub-algebra,
\end{enumerate}
play an important role in further analysis of the structure of the CC equations and 
we will refer all  sub-algebras satisfying requirements (1) and (2)  as sub-system embedding sub-algebras (SESs).

In Ref.~\cite{safkk} it has been shown that $\mathfrak{g}^{(N)}(1_{\rm R},y_{\rm S})$
along with $\mathfrak{g}^{(N)}(x_{\rm R},1_{\rm S})$ form SESs for CCSD and 
$\mathfrak{g}^{(N)}(2_{\rm R},y_{\rm S})$ along with $\mathfrak{g}^{(N)}(x_{\rm R},2_{\rm S})$ 
are SESs for CCSDTQ. The largest sub-algebras in these classes are 
$\mathfrak{g}^{(N)}(1_{\rm R})$ and $\mathfrak{g}^{(N)}(2_{\rm R})$ sub-algebras. In 
Ref.~\cite{safkk}, we have also demonstrated that for any SES $\mathfrak{h}$ 
(corresponding to some CC approximation) the standard 
\begin{eqnarray}
	Q_{\rm int} \bar{H}_N|\Phi\rangle &=& 0  \;,\label{rr12} \\
	Q_{\rm ext} \bar{H}_N |\Phi\rangle &=& 0 \;, \label{rr34} \\
	\langle\Phi|\bar{H}|\Phi\rangle &=& E \;, 
\label{enest}
\end{eqnarray}
and hybrid 
\begin{eqnarray}
	(P+Q_{\rm int}) \bar{H}_{\rm ext} e^{T_{\rm int}}|\Phi\rangle &=& E(P+Q_{\rm int}) 
    e^{T_{\rm int}}|\Phi\rangle \;, \label{pp12} \\
	Q_{\rm ext} \bar{H}_N |\Phi\rangle &=& 0 \;, 
\label{pp34} 
\end{eqnarray}
representations of CC equations are equivalent at the solution.
In Eqs. (\ref{rr12})-(\ref{pp34}) we used simplified notations: 
(1) the projection operators
	$Q_{\rm int}(\mathfrak{h})$ and $Q_{\rm ext}(\mathfrak{h})$ 
	($Q= Q_{\rm int}(\mathfrak{h})+ Q_{\rm ext}(\mathfrak{h})$)
	project onto subspaces spanned by all excited configurations 
	generated by acting $T_{\rm int}(\mathfrak{h})$ 
	and $T_{\rm ext} (\mathfrak{h})$ onto the reference function, respectively, 
(2) $\bar{H}_{N}=\bar{H}-\langle\Phi|\bar{H}|\Phi\rangle$ is the normal
	product form of the similarity transformed Hamiltonian $\bar{H}$, 
(3) the $\bar{H}_{\rm ext} $ operator is defined as 
	$\bar{H}_{\rm ext} \equiv \bar{H}_{\rm ext}(\mathfrak{h}) = e^{-T_{\rm 
    ext}(\mathfrak{h})}He^{T_{\rm ext}(\mathfrak{h})}$, and 
(4) for notational simplicity we used the following notational convention,
	\begin{eqnarray}
		T_{\rm int} &\equiv& T_{\rm int}(\mathfrak{h}) \;, \label{simp1} \\
		T_{\rm ext} &\equiv& T_{\rm ext}(\mathfrak{h}) \;, \label{simp2} \\
		Q_{\rm int} &\equiv& Q_{\rm int}(\mathfrak{h}) \;, \label{simp3} \\
		Q_{\rm ext} &\equiv& Q_{\rm ext}(\mathfrak{h}) \;. \label{simp4}
	\end{eqnarray}
The above mentioned equivalence means that  cluster amplitudes corresponding to 
excitations included in $\mathfrak{h}$ can be obtained in a diagonalization procedure. 
Moreover, the standard form of the CC energy expression (given by Eq. (\ref{ccene})) is a  
special case of Eq. (\ref{pp12}) corresponding to $\mathfrak{h}=\mathfrak{g}^{(N)}(0)$ 
(where $\mathfrak{g}^{(N)}(0)$ contains no excitations) - in this case 
$T_{\rm int}(\mathfrak{h})=0$.

An immediate consequence of the above equivalence is the fact that the energy of the 
entire systems can be obtained at the solution as an eigenvalue of the effective 
Hamiltonian operator $\bar{H}_{\rm ext}^{\rm eff}(\mathfrak{h})$,
\begin{equation}
	\bar{H}_{\rm ext}^{\rm eff}(\mathfrak{h})=(P+Q_{\rm int}(\mathfrak{h})) 
    \bar{H}_{\rm ext}(\mathfrak{h}) (P+Q_{\rm int}(\mathfrak{h})) \;,
\label{extdiag}
\end{equation}
is the corresponding complete active space. 
By construction, the cluster amplitudes $T_{\rm ext}$, used to define 
$\bar{H}_{\rm ext}(\mathfrak{h})$, are decoupled from cluster amplitudes $T_{\rm int}$ that define the components of the corresponding eigenvector. 
One should also notice that the  many-body expansion of 
$\bar{H}_{\rm ext}(\mathfrak{h})$
may contain effective interactions involving higher-than-pairwise interactions.

\begin{figure}
	\includegraphics[angle=0, width=0.55\textwidth]{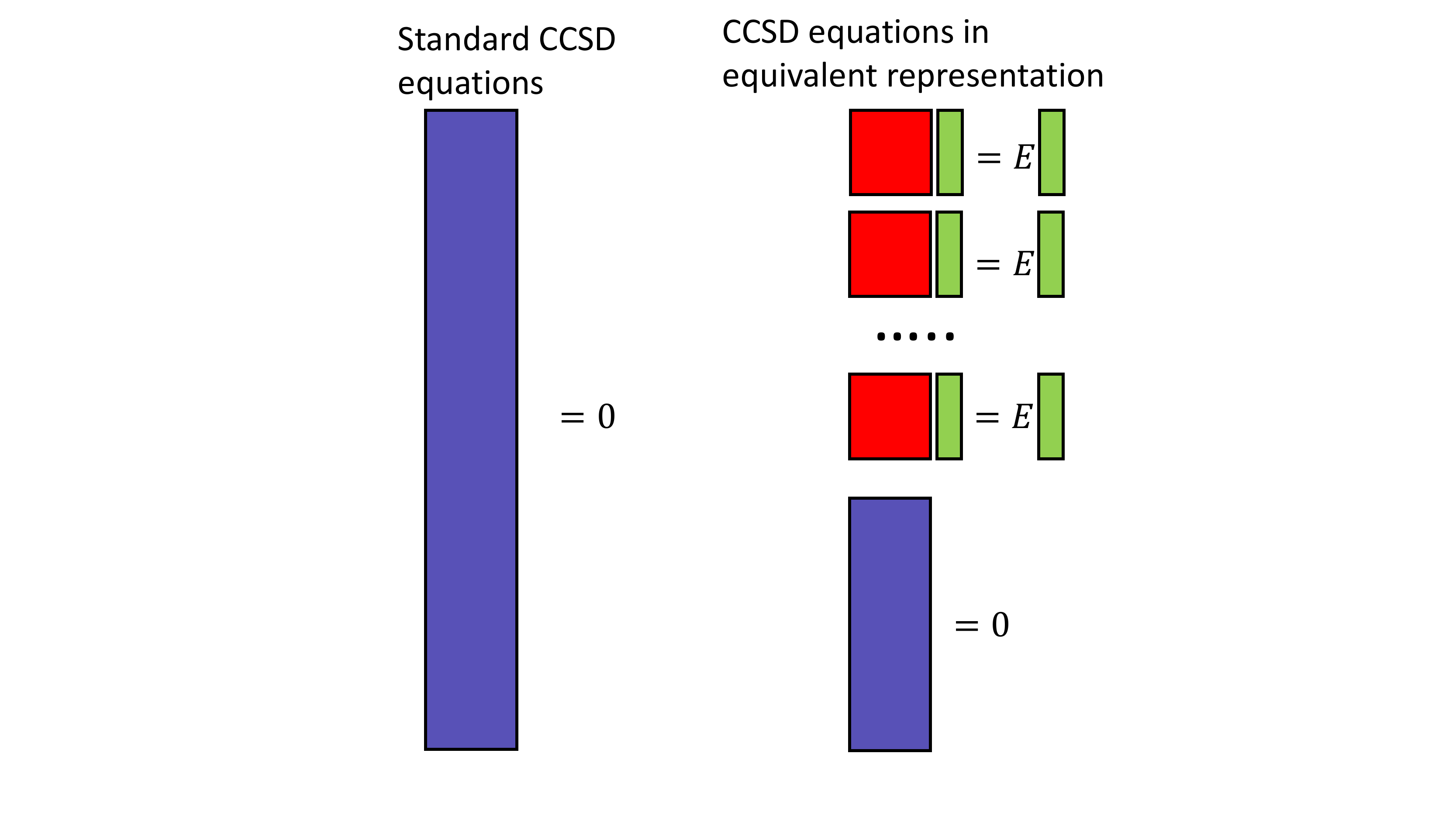}
	\caption{Two equivalent representations of the CCSD equations. The left part 	
    schematically designates their standard form $Q(He^{T_1+T_2})_C|\Phi\rangle$ (blue 
    block) while the right part (based on the utilization of various subsystems 
    embedding sub-algebras) contains several coupled eigenvalue problems corresponding 
    to various SESs and projections of $(He^{T_1+T_2})_C|\Phi\rangle$ on configurations 
    not included in a corresponding set of SESs 
    (i.e., $Q_{\rm ext} \bar{H}_N|\Phi\rangle$ symbolically designated  by the blue block). }
\label{fig1}
\end{figure}

\begin{figure}
	\includegraphics[angle=0, width=0.5\textwidth]{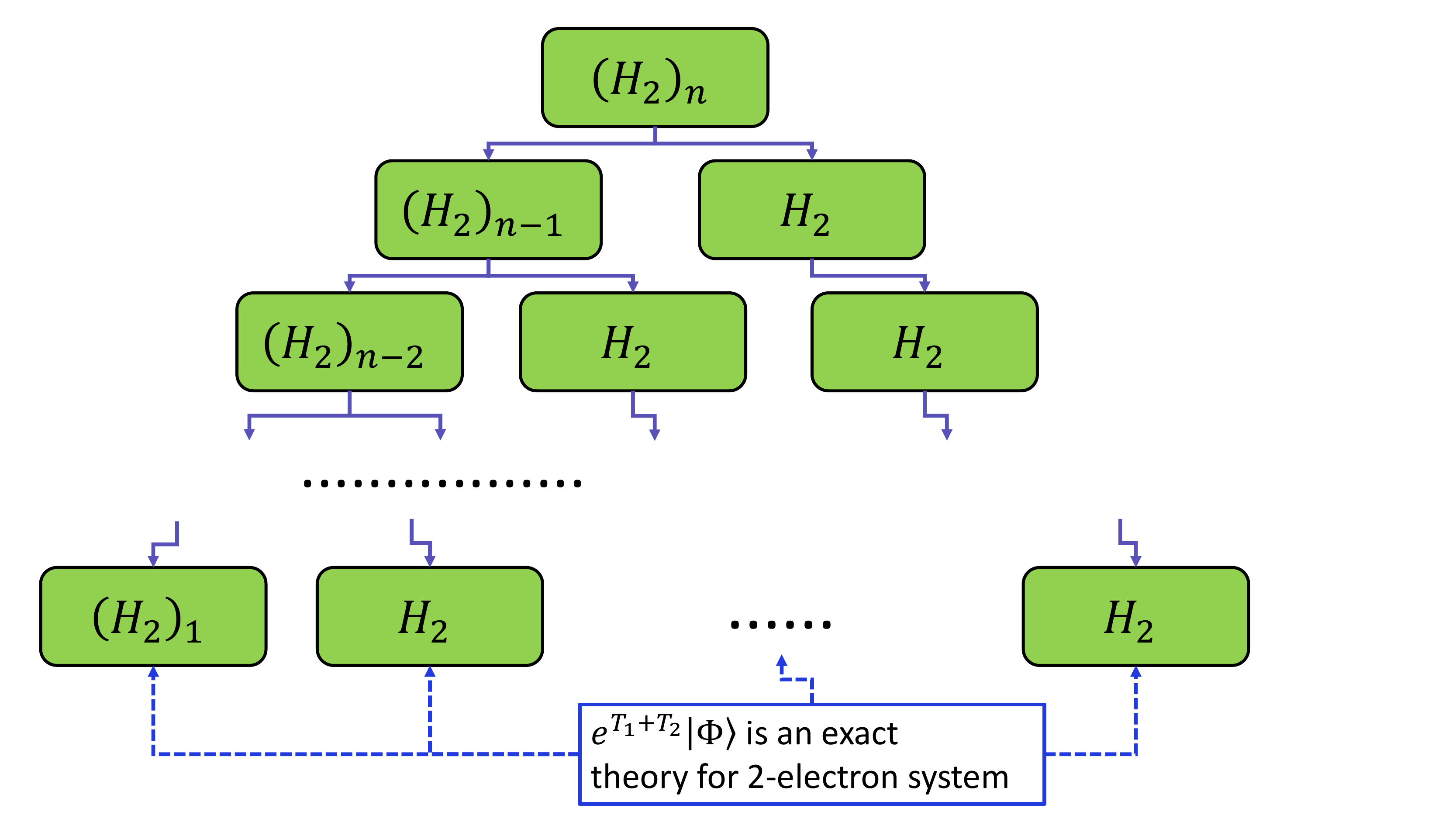}
	\caption{A schematic representation of the system partitioning 
    (assembly of $n$ interacting H$_2$ molecules, (H$_2$)$_n$)
    into non-interacting 
    two-electron sub-systems (non-interacting H$_2$ molecules). 
    For the exact CC formalism, for each H$_2$ sub-system there exist SES of the 
    $\mathfrak{g}^{(N)}(1_{\rm R})$ type that provides excitations needed to describe  
    H$_2$ molecule at the non-interacting sub-system limit. For simplicity, we assume      
    that all molecular orbitals describing  (H$_2$)$_n$ evolve in the non-interacting subsystem limit into orbitals localized on the non-interacting H$_2$ systems.
	}
\label{fig0}
\end{figure}

\begin{figure}
	\includegraphics[angle=0, width=0.55\textwidth]{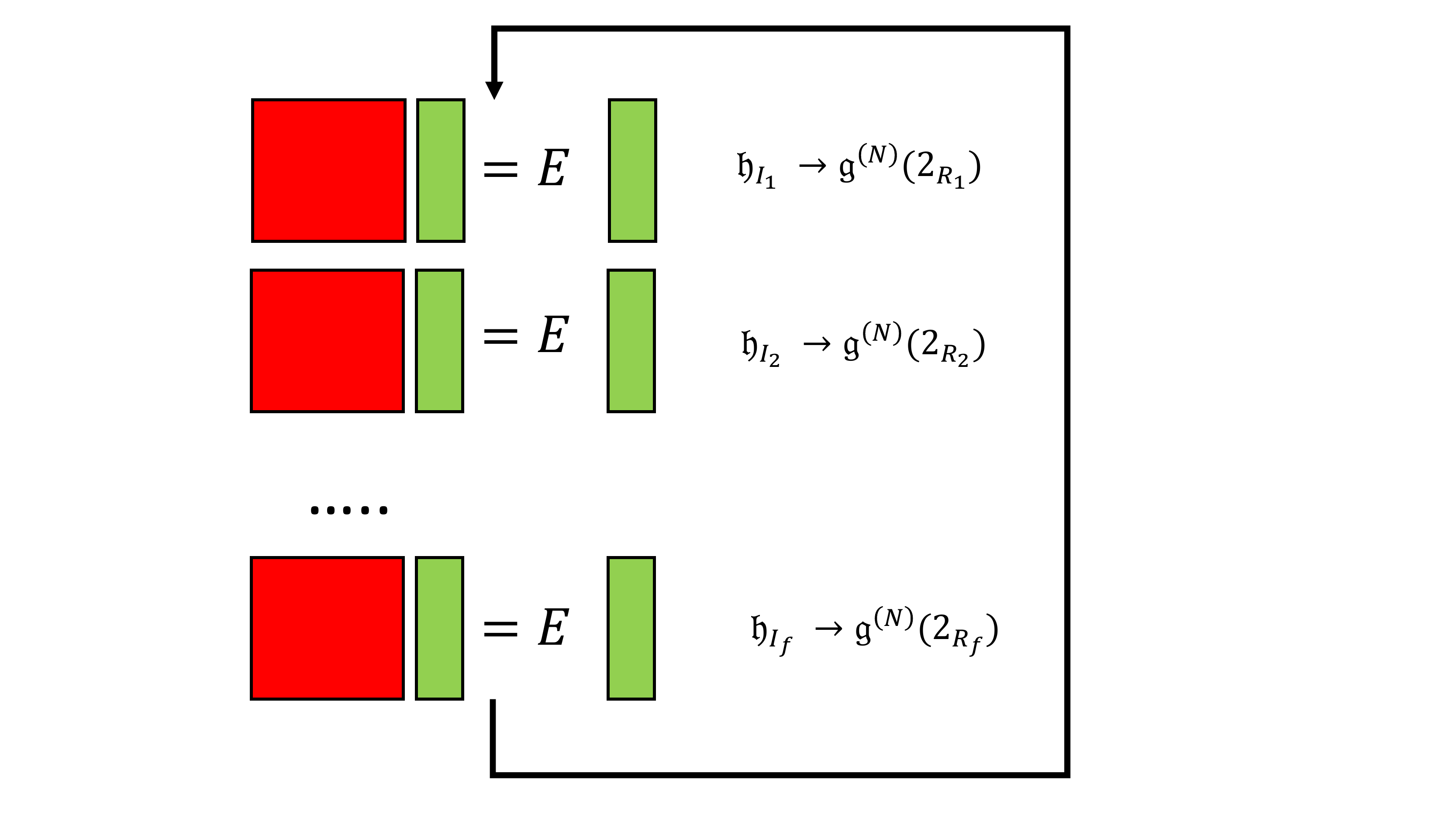}
	\caption{Form of the CC equations (SAF-CCSD(2) formalism of Ref.~\cite{safkk}) based 
    on the inclusion of excitations from all $\mathfrak{g}^{(N)}(2_{\rm R})$ sub-algebras.
	Each eigenvalue problem corresponds to a specific effective Hamiltonian 
    $\bar{H}_{\rm ext}^{\rm eff}(\mathfrak{h})$ where $\mathfrak{h}$
    is  of the $\mathfrak{g}^{(N)}
    (2_{\rm R})$ type 
     ($\mathfrak{h}=\mathfrak{g}^{(N)}(2_{\rm R_1}),\ldots , 
       \mathfrak{g}^{(N)}(2_{\rm R_f}$)).}
\label{fig2}
\end{figure}

Properties of SESs induced eigenvalue problems can also be utilized to define alternative ways of forming CC approximations and corresponding CC equations.
For example, the CCSD equations can be re-cast in the form shown in Fig. \ref{fig1}. 
The form of the decomposition shown in Fig. \ref{fig1} can also be viewed as an "echo" of 
the fact that CCSD theory is an exact theory for subsystem decomposed into 
non-interacting two-electron systems (an example is shown in Fig. \ref{fig0}).
It is also interesting to notice that for the exact CC theory for closed-shell systems discussed here, there exist a chain of various 
types of SESs that meet requirements (1) and (2). For instance, for the exact formulation one can consider a chain of 
SESs defined as 
\begin{equation}
	\mathfrak{g}^{(N)} \rightarrow \ldots \mathfrak{g}^{(N)} (i_{\rm R}) \rightarrow \ldots 
    \mathfrak{g}^{(N)} (2_{\rm R}) \rightarrow \mathfrak{g}^{(N)} (1_{\rm R}) \;,
\label{chain1}
\end{equation}
which results in separate eigenvalue problems corresponding to sub-systems of various sizes.
This observation can be used to define a new CC approximations
where instead of referring to adding higher and higher ranks 
of excitations as a design principle (used to defined standard approximations such as CCD, 
CCSD, CCSDT, CCSDTQ, etc.) one can envision a strategy based on the inclusion excitations in the 
cluster operator that belong to a specific class (or classes) of SESs. For example, in 
Ref.~\cite{safkk} we discussed  an approximation (sub-algebra flow CCSD(2) approximation 
(SAF-CCSD(2)) that employs all amplitudes contained in all $\mathfrak{g}^{(N)}(2_{\rm R})$ SESs, which  leads to a CC model that contains all singly and double excited cluster amplitudes 
and selected subsets of triply and quadruply excited ones and where the CC equations can 
be represented as a conglomerate of coupled eigenvalue problems shown in Fig. \ref{fig2} 
(for more detailed discussion see Ref.~\cite{safkk}).
It can also be shown that selecting cluster amplitudes based on subsystem embedding algebras 
is a natural way of introducing a notion of seniority number intensively studied in the 
context of CC applications with strongly correlated systems.

To summarize this section,  techniques based on the utilization of the sub-system 
embedding sub-algebras can be used to downfold the full electronic Hamiltonian to an
arbitrary active space corresponding to some subsystem embedding sub-algebra. For 
example, using SES one could downfold electronic Hamiltonian to the active space 
(usually containing highest occupied and lowest unoccupied orbitals) that contains the 
most important contributions to the electronic wave function of interest. 
This fact may be especially appealing in the context of 
quantum  algorithms such as quantum phase estimation (QPE)~\cite{luis1996optimum,
cleve1998quantum,berry2007efficient,childs2010relationship,wecker2015progress,
haner2016high,poulin2017fast} or variational quantum eigensolver 
(VQE)~\cite{peruzzo2014variational,mcclean2016theory,openfermion,fontalvo2017strategies,
PhysRevA.95.020501,kandala2018extending,PhysRevX.8.011021}. 
Unfortunately, the main 
caveat is related to the fact that effective Hamiltonians 
$\bar{H}_{\rm ext}(\mathfrak{h)}$ are non-Hermitian, which is a major obstacle in using these algorithms. The main goal of the following section is to redefine 
unitary CC formalism to provide efficient downfolding algorithm, which  yields Hermitian 
effective Hamiltonian(s) and at the same time assures the same properties as single-reference CC effective Hamiltonians $\bar{H}_{\rm ext}(\mathfrak{h)}$ discussed in this 
section. 


\section{Unitary CC method}
\label{uccsection}

The unitary CC (UCC) formulations have been introduced in Refs.~\cite{pal1984use,
hoffmann1988unitary,unitary1,kutzelnigg1991error,unitary2,sur2008relativistic,evangelista2011alternative,
cooper2010benchmark,harsha2018difference}
 and have been intensively studied in the recent years in the 
context of various quantum algorithms~\cite{peruzzo2014variational,mcclean2016theory,romero2018strategies,shen2017quantum,motta2018low}.

The standard UCC Ansatz (the generalization of standard UCC Ansatz has recently been discussed in the literature, for details see Ref.~\cite{lee2018generalized})
\begin{equation}
	|\Psi\rangle=e^{\sigma} |\Phi\rangle \;
\label{ucc1}
\end{equation}
is reminiscent of standard single-reference expansion (\ref{ccp}) with the difference 
that the $\sigma$ operator is anti-Hermitian, i.e.,
\begin{equation}
	\sigma^{\dagger} = - \sigma \;.
\label{sigma1}
\end{equation}
This property of the $\sigma$ operator leads to a non-terminating character of many-body expansion 
for the wave function (\ref{ucc1}). The anti-Hermitian character of $\sigma$ can be 
assured by the specific form of the $\sigma$ operator, which in most formulations is represented as 
\begin{equation}
	\sigma = T-T^{\dagger} \;,
\label{sigma2}
\end{equation}
where $T$ has exactly the same structure as discussed in Eq. (\ref{tcomp}). In this and 
following Sections we will focus on exact (i.e. including all possible excitations) 
formulation of the UCC theory. 
%

In the analysis of the UCC formalisms we will refer to two standard formulas for 
operator exponentials:\\
\noindent 1.) The Baker-Campbell-Hausdorff (BCH) formula:
\begin{equation}
	e^X e^Y = e^{X+Y+\sum_{k=1}^{\infty} C_{\rm BCH}^{(k)}(X,Y)}\;,
\label{bch1}
\end{equation}
where commutators $C_{\rm BCH}^{(k)}$ are defined as $C_{\rm BCH}^{(1)}(X,Y)=\frac{1}{2}[X,Y]$, 
$C_{\rm BCH}^{(2)}(X,Y)=\frac{1}{12}([X,[X,Y]]+[Y,[Y,X]])$, etc. 

\noindent 2.) The transposed variant of the Zassenhaus formula~\cite{wilcox1967exponential,scholz2006note}: 
\begin{eqnarray}
	e^{X+Y}&=& \lbrace \prod_{\infty}^{k=2} e^{C_{\rm Z}^{(k)}(X,Y)} \rbrace e^Y e^X \nonumber \\
	&=& R_{\rm Z}(X,Y) e^Y e^X \;,
\label{zass1}
\end{eqnarray}
where $C_{\rm Z}^{(2)}=\frac{1}{2}[X,Y]$, $C_{\rm Z}^{(3)}=\frac{1}{3}[Y,[[X,Y]] +\frac{1}{6}[X,[X,Y]]$, etc. 
The $R_{\rm Z}(X,Y)$ function is defined as 
\begin{equation}
	R_{\rm Z}(X,Y)=\prod_{\infty}^{k=2} e^{C_{\rm Z}^{(k)}(X,Y)}\;, 
\label{rzf1}
\end{equation}
where its inverse is given by the formula 
\begin{equation}
	R_{\rm Z}^{-1}(X,Y)=\prod_{k=2}^{\infty} e^{-C_{\rm Z}^{(k)}(X,Y)}\;.
\label{rzf2}
\end{equation}

The equations for $\sigma$ operator in the exact limit can be obtained by substituting 
ansatz (\ref{ucc1}) into the Schr\"odinger equation
\begin{equation}
	He^{\sigma}|\Phi\rangle = E e^{\sigma} |\Phi\rangle \;.
\label{schucc}
\end{equation}
The equivalent representation, which  provides the explicitly connected form of the equations, 
can be obtained by multiplying both sides by $e^{-\sigma}$ operator and decoupling 
equations for amplitudes from the equation for energy (see Ref.~\cite{unitary2,evangelista2011a} for details), 
i.e.,
\begin{eqnarray}
	Qe^{-\sigma}He^{\sigma}|\Phi\rangle &=& 0 \label{ucceq1} \\
	\langle\Phi|e^{-\sigma}He^{\sigma}|\Phi\rangle &=& E \label{ucceq2} \;.
\end{eqnarray}

In analogy to the standard single-reference approach let us explore if partitioning of 
$\sigma$ into part belonging to some SES $\mathfrak{h}$
($\sigma_{\rm int}\equiv \sigma_{\rm int}(\mathfrak{h})$) and its complement 
($\sigma_{\rm ext}\equiv \sigma_{\rm ext}(\mathfrak{h})$) leads to a downfolding of 
Hamiltonian in the form discussed in Section \ref{sessec}.
For this purpose we will (1) apply Zassenhaus formula (\ref{zass1}) to factorize 
$e^{\sigma_{\rm int}+\sigma_{\rm ext}}$, (2) premultiply (\ref{schucc}) from the left by 
$e^{-\sigma_{\rm ext}} R_Z^{-1}(\sigma_{\rm int},\sigma_{\rm ext})$, and (3) project 
resulting equations onto $P+Q_{\rm int}$ space. This procedure leads to the following 
equations 
\begin{align}
	 &\Big((P+Q_{\rm int}) e^{-\sigma_{\rm ext}} R_Z^{-1}(\sigma_{\rm int},\sigma_{\rm ext}) H
	  R_Z(\sigma_{\rm int},\sigma_{\rm ext})\\\nonumber
	&e^{\sigma_{\rm ext}} e^{\sigma_{\rm int}}\Big)|\Phi\rangle
	= E (P+Q_{\rm int}) e^{\sigma_{\rm int}}|\Phi\rangle \;.
\label{eqm1}
\end{align}
%
%
For simplicity, the above equation can be re-written as 
\begin{equation}
	(P+Q_{\rm int}) \bar{H}_{\rm ext}^{\rm UCC} e^{\sigma_{\rm int}}|\Phi\rangle
	=E (P+Q_{\rm int}) e^{\sigma_{\rm int}}|\Phi\rangle
\label{ucceig1}
\end{equation}
where the $\bar{H}_{\rm ext}^{\rm UCC}$ operator is defined as 
\begin{equation}
	\bar{H}_{\rm ext}^{\rm UCC}=
	e^{-\sigma_{\rm ext}} R_Z^{-1}(\sigma_{\rm int},\sigma_{\rm ext}) H R_Z(\sigma_{\rm int},\sigma_{\rm ext})
	e^{\sigma_{\rm ext}} \;,
\label{ucchash}
\end{equation}
or equivalently
\begin{equation} 
	\bar{H}_{\rm ext}^{\rm eff(UCC)}e^{\sigma_{\rm int}}|\Phi\rangle = E e^{\sigma_{\rm int}}|\Phi\rangle\;,
\label{ucchash2}
\end{equation}
where the effective Hamiltonian $\bar{H}_{\rm ext}^{\rm eff(UCC)}$ in 
Eq. (\ref{ucchash2}) is defined as 
\begin{align}
	\bar{H}_{\rm ext}^{\rm eff(UCC)}=&
	\Big((P+Q_{\rm int}) e^{-\sigma_{\rm ext}} R_Z^{-1}(\sigma_{\rm int},\sigma_{\rm ext}) H\\\nonumber &R_Z(\sigma_{\rm int},\sigma_{\rm ext})
	e^{\sigma_{\rm ext}} (P+Q_{\rm int})\Big).
\label{uccheff}
\end{align}
In the above equation we use the fact that 
\begin{equation}
	e^{\sigma_{\rm int}}|\Phi\rangle = (P+Q_{\rm int}) e^{\sigma_{\rm int}}|\Phi\rangle \;.
\label{uccpqa}
\end{equation}
It can be shown that the form of the active-space eigenvalue-type problem 
(\ref{ucchash2}) is equivalent to the $(P+Q_{\rm int})$ projections of the connected 
form of the UCC equations described by Eq. (\ref{ucceq1}). This can be shown by 
introducing the resolution of identity $e^{-\sigma_{\rm int}} e^{\sigma_{\rm int}}$ to 
the left of $\bar{H}_{\rm ext}^{\rm UCC}$ in Eq. (\ref{ucceig1}) and noticing that 
$e^{-\sigma_{\rm int}} \bar{H}_{\rm ext}^{\rm UCC} e^{\sigma_{\rm int}}
=e^{-\sigma}H e^{\sigma}$. Then the resulting equation takes the form 
\begin{align}
	&(P+Q_{\rm int}) e^{\sigma_{\rm int}} (P+Q_{\rm int}) e^{-\sigma}He^{\sigma}|
    \Phi\rangle 
    \nonumber\\
    &\quad= E (P+Q_{\rm int})e^{\sigma_{\rm int}} |\Phi\rangle.
\label{analysis1}
\end{align}
Using matrix representation of the $\sigma_{\rm int}$ operator in the CAS space denoted 
as $\bm{\sigma}_{\rm int}$ this equation can be re-written as 
\begin{equation}
	[e^{\bm{\sigma}_{\rm int}}] [\bm{x}] = 0 \;,
\label{llineq}
\end{equation} 
where the first component of vector $\bm{x}$ corresponds to $\langle\Phi|e^{-
\sigma}He^{\sigma}|\Phi\rangle -E$ while the remaining components correspond to 
projections of $e^{-\sigma}He^{\sigma}|\Phi\rangle$ onto excited (with respect to the 
reference determinant $|\Phi\rangle$) configurations $\langle\Phi_{\Delta}^{\rm (CAS)}|$ 
belonging to $\mathfrak{h}$-induced CAS. The $[e^{\bm{\sigma}_{\rm int}}]$ matrix is 
also non-singular, which is a consequence of the algebraic structure of $\sigma_{\rm int}$ 
rather than a particular values of cluster amplitudes defining $T_{\rm int}$ and $T_{\rm 
int}^{\dagger}$. To see it let us calculate its determinant
\begin{equation}
	{\rm det}(e^{\bm{\sigma}_{\rm int}})=e^{{\rm Tr}(\bm{\sigma}_{\rm int})} 
\end{equation}
where $\bm{T}_{\rm int}$ and $\bm{T}^{\dagger}_{\rm int}$ are matrix representations of 
$T_{\rm int}$ and $T^{\dagger}_{\rm int}$ in CAS. Since $T_{\rm int}$ and 
$T^{\dagger}_{\rm int}$ contains either excitations or de-excitations we have 
\begin{equation}
	{\rm Tr}(\bm{\sigma}_{\rm int})={\rm Tr}(\bm{T}_{\rm int}-\bm{T}^{\dagger}_{\rm int})=0 \;,
\end{equation}
which means that ${\rm det}(e^{\bm{\sigma}_{\rm int}})=1$ and $(e^{\bm{\sigma}_{\rm 
int}})$ is a non-singular operator. Therefore, the only solution of Eq. (\ref{llineq}) 
corresponds to $\bm{x}=0$, which proves the equivalence of Eq. (\ref{ucceig1}) with $P$ 
and $Q_{\rm int}$ projections of Eqs. (\ref{ucceq1}) and (\ref{ucceq2}).

Although $\bar{H}_{\rm ext}^{\rm UCC}$ ($\bar{H}_{\rm ext}^{\rm eff(UCC)}$) is 
Hermitian, in contrast to $\bar{H}_{\rm ext}$ from Eq. (\ref{pp12}) the 
$\bar{H}_{\rm ext}^{\rm UCC}$ ($\bar{H}_{\rm ext}^{\rm eff}$) operator does not decouple 
$\sigma_{\rm ext}$ amplitudes from the $\sigma_{\rm int}$ ones. If $\mathfrak{h}$ is 
chosen to contain highest/lowest lying occupied/virtual orbitals we can view this 
property of $\bar{H}_{\rm ext}^{\rm UCC}$ as mixing low- and high-energy components of 
the wave function (or using quantum chemical lingua, as a mixing of static and dynamic 
correlation effects). In the next section we will discuss how to re-instate the 
separation of effective Hamiltonian while maintaining its Hermitian character. This will 
have important consequences on how the approximate formulations can be constructed.



\section{Alternative UCC expansions}

In this section we will discuss properties of a  UCC Ansatz given by the product 
of two unitary transformations (double unitary CC (DUCC) expansion or, for the reasons 
explained later, tailored unitary CC formulation) which explicitly employs  the 
partitioning induced by some SES $\mathfrak{h}$
\begin{equation}
	|\Psi\rangle=e^{\sigma_{\rm ext}} e^{\sigma_{\rm int}}|\Phi\rangle \;.
\label{ducc1}
\end{equation}
The above expansion is driven by similar ideas as a class of CC methods that utilize 
double similarity transformations and have been extensively discussed in the 
literature~\cite{stolarczyk1985coupled,meissner1995effective,nooijen1996many,
nooijen1997new,meissner1998fock}. Moreover, ansatz (\ref{ducc1}) can also be viewed as a 
unitary generalization of the tailored CC formalism (TCC)~\cite{tailored1,tailored2,
melnichuk12,melnichuk14}, where equations for $\sigma_{\rm int}$ are represented in the 
form of eigenvalue problem discussed in previous sections. 

In analogy to the previous section,  we will focus on the exact 
formulation of the DUCC approach that includes all possible excitations in the $\sigma_{\rm 
ext}$ and $\sigma_{\rm int}$ operators. Since using BCH expansion 
(\ref{bch1})
the DUCC expansion can be transformed to the alternative single-reference ansatz
\begin{equation}
	e^{\sigma_{\rm ext}} e^{\sigma_{\rm int}}|\Phi\rangle = e^D |\Phi\rangle\;,
\label{dequ}
\end{equation}
where $D$ is anti-Hermitian ($D^{\dagger}=-D$), the DUCC formalism can also be viewed 
as a special case of a unitary CC ansatz.

In analogy to the UCC formulation, when the double UCC ansatz 
(\ref{ducc1}) is introduced into Schr\"odinger equation
\begin{equation}
	He^{\sigma_{\rm ext}} e^{\sigma_{\rm int}}|\Phi\rangle= E e^{\sigma_{\rm ext}} e^{\sigma_{\rm int}}|\Phi\rangle
\label{uccsch}
\end{equation}
it can be rewritten in the equivalent form which 
decouples equations for cluster amplitudes from the equation for energy
\begin{eqnarray}
	Qe^{-\sigma_{\rm int}}e^{-\sigma_{\rm ext}} H e^{\sigma_{\rm int}}e^{\sigma_{\rm ext}} |\Phi\rangle &=& 0 \;,
\label{uccd2eq} \\
	\langle\Phi|e^{-\sigma_{\rm int}}e^{-\sigma_{\rm ext}} H e^{\sigma_{\rm int}}e^{\sigma_{\rm ext}} |\Phi\rangle &=& E \;.
\label{uccd2ene}
\end{eqnarray}
One can show that the equations (\ref{uccsch}) corresponding to projections onto $(P+Q_{\rm ext})$ sub-space
can be written in equivalent form as an eigenvalue problem 
\begin{equation}
	(P+Q_{\rm int}) \bar{H}_{\rm ext}^{\rm DUCC} e^{\sigma_{\rm int}} |\Phi\rangle = 
    E(P+Q_{\rm int}) e^{\sigma_{\rm int}}|\Phi\rangle
\label{duccstep1}
\end{equation}
or equivalently, using effective Hamiltonian language,
\begin{equation}
	\bar{H}_{\rm ext}^{\rm eff(DUCC)} e^{\sigma_{\rm int}} |\Phi\rangle = Ee^{\sigma_{\rm int}}|\Phi\rangle
\label{duccstep2}
\end{equation}
where
\begin{equation}
	\bar{H}_{\rm ext}^{\rm eff(DUCC)} = (P+Q_{\rm int}) \bar{H}_{\rm ext}^{\rm DUCC} (P+Q_{\rm int})
\label{equivducc}
\end{equation}
and 
\begin{equation}
	\bar{H}_{\rm ext}^{\rm DUCC} =e^{-\sigma_{\rm ext}}H e^{\sigma_{\rm ext}} \;.
\label{duccexth}
\end{equation}
To show this fact it suffices to introduce the resolution of identity $e^{\sigma_{\rm 
int}}e^{-\sigma_{\rm int}}$ to the left of the $\bar{H}_{\rm ext}^{\rm DUCC} $ operator 
in Eq. (\ref{duccstep1}) and notice that $e^{-\sigma_{\rm int}}\bar{H}_{\rm ext}^{\rm 
DUCC} e^{\sigma_{\rm int}}=e^{-\sigma_{\rm int}}e^{-\sigma_{\rm ext}} H e^{\sigma_{\rm 
int}}e^{\sigma_{\rm ext}}$. Next, in analogy to Eqs. (\ref{analysis1}) and 
(\ref{llineq}), Eq. (\ref{equivducc}) can be represented as 
\begin{equation}
	[e^{\bm{\sigma}_{\rm int}}] [\bm{y}] = 0 \;,
\label{llineq2}
\end{equation}
where the first component of the $[\bm{y}]$ vector is equivalent to 
$\langle\Phi|e^{-\sigma_{\rm int}}e^{-\sigma_{\rm ext}} H 
e^{\sigma_{\rm int}}e^{\sigma_{\rm ext}} |\Phi\rangle-E$
while the remaining components correspond to projections of 
$e^{-\sigma_{\rm int}}e^{-\sigma_{\rm ext}} H 
e^{\sigma_{\rm int}}e^{\sigma_{\rm ext}} |\Phi\rangle$
onto excited configurations belonging to $Q_{\rm int}$. Given the non-singular character 
of the $[e^{\bm{\sigma}_{\rm int}}]$ matrix, this proves the equivalence of these two 
representations. 

By construction, the DUCC effective Hamiltonian is Hermitian and in contrast to the 
UCC case it is expressible in terms of the external $\sigma_{\rm ext}$ amplitudes only,  
providing in this way in  Eq. (\ref{duccstep2}) a rigorous decoupling of degrees of freedom 
corresponding to ${\sigma_{\rm ext}}$ and ${\sigma_{\rm int}}$ in the sense of 
discussion of Section \ref{sessec}. Since amplitudes defining 
$\sigma_{\rm ext}$ are 
characterized by larger perturbative denominator compared to the $\sigma_{\rm int}$, 
where small denominators may occur, it is much safer to determine
$\sigma_{\rm ext}$ using 
perturbative techniques. The Eq. (\ref{duccstep2}) also offers  a possibility of 
downfolding the Hamiltonian to the $(P+Q_{\rm int})$ space where the correlation effects 
from $Q_{\rm ext}$ can be included through the $\sigma_{\rm ext}$ operator, which makes 
calculations with the downfolded Hamiltonian amenable for quantum computing even for larger systems. 

In the following part of the paper 
we will discuss an approximate form of the second 
quantized representation of the $\bar{H}_{\rm ext}^{\rm eff(DUCC)}$ operator
(denoted as the $\Gamma$ operator)
for a specific choice of SES containing all occupied 
and lowest-lying virtual spin orbitals to define the SES active space (i.e. all 
occupied indices $i,j,\ldots$ and some small subset of virtual spin orbitals 
$a,b,\ldots$ are deemed active). In this case amplitudes defining the $\sigma_{\rm 
ext}$ operator must carry at least one inactive virtual orbital. One can view this 
procedure as a downfolding of an essential part of the virtual spin-orbital space.
This process will 
consist of two steps: (1) expansion of $\bar{H}_{\rm ext}^{\rm eff(DUCC)}$  in powers of $\sigma_{\rm ext}$ operator and (2) 
approximation of $\sigma_{\rm ext}$ operator. 
The 
$\Gamma$ operator is defined by strings of creation/annihilation operators that carry 
only active spin orbitals and can be written as
\begin{equation}
	\Gamma=(\bar{H}_{\rm ext}^{\rm DUCC})_{\rm act} \;,
\label{xi1}
\end{equation}
where subscript ${\rm act} $ designates terms of a given operator expression that contains 
creation/annihilation operators carrying active-space spin-orbital labels. Using Baker-Campbell-Hausdorff formula one can re-cast the above equation in the form of infinite 
expansion:
\begin{equation}
	\Gamma=(H)_{\rm act} +([H,\sigma_{\rm ext}])_{\rm act} + \frac{1}{2!}([[H,\sigma_{\rm ext}],\sigma_{\rm ext}])_{\rm act} + \ldots
\label{xiex}
\end{equation}
Using particle-hole (ph) formalism one can also expand the $\Gamma$ operator into the sum of its many-body components $\Gamma_i$
\begin{equation}
\Gamma=\Gamma_{\rm scalar} + \Gamma_1+ \Gamma_2 + \Gamma_3 + \ldots  + \Gamma_{N_{e,{\rm act}}} \;,
\label{gammb}
\end{equation}
where $N_{e,{\rm act}}$ designates the number of active electrons and
\begin{eqnarray}
\Gamma_1 &=& \sum_{PQ} \gamma^P_Q N[a_P^{\dagger} a_Q]^{\rm ph}   , \label{g1ga} \\\
\Gamma_2 &=& \frac{1}{(2!)^2}\sum_{PQRS} \gamma^{PQ}_{RS} N[a_P^{\dagger} a_Q^{\dagger} a_S a_R ]^{\rm ph}, \label{g2ga} \\
\vdots && \nonumber \\
\Gamma_i &=& 
\frac{1}{(i!)^2}
\sum_{\substack{P_1,\ldots,P_i \\ Q_1,\ldots,Q_i}}
\gamma^{P_1\ldots P_i}_{Q_1\ldots Q_i} N[a_{P_1}^{\dagger} \ldots a_{P_i}^{\dagger} a_{Q_i} \ldots a_{Q_1}]^{\rm ph}, \;\;\;\;\;\;  \label{giga} \\
\vdots && \nonumber
\end{eqnarray}
where $N[...]^{\rm ph} $ designates the particle-hole normal product form of a given operator expression,
$P,Q,R,S,\ldots $ designate 
general active spin-orbital indices, and $\Gamma_{\rm scalar}$ designates full contracted (scalar) part of the $\Gamma$ operator with respect to the 
particle-hole  vacuum $|\Phi\rangle$. In the above expansion we also assume that all multidimensional tensors $\gamma^{P_1\ldots P_i}_{Q_1\ldots Q_i}$
are antisymmetric with respect to the permutations among the sets of lower and upper spin orbital indices.

%
In practical realizations of the DUCC formalism one has to truncate both the many-body 
$\Gamma$ expansion given by Eq.(\ref{giga}) as well as excitation level included in the $\sigma_{\rm ext}$ operator. 
Below, as a specific example  we describe a variant of $\Gamma$-operator approximations based on the inclusion of one- and 
two-body interactions/excitations. This will also illustrate the benefits of  using  a hybrid particle-hole and physical vacuum 
second-quantized representations of 
all operators involved in the approximation.
Following similar ideas as discussed in Ref.~\cite{unitary1}, where energy functionals 
were constructed based on the order of the  energy contributions, one can select terms in 
expansion (\ref{xiex}) based on the perturbative analysis of contributing terms.
For  example, using particle-hole formalism, the normal product form of the 
electronic Hamiltonian $H_N$ ($H_N=H-\langle\Phi|H|\Phi\rangle$)
 can be split  into its one-particle part 
$F_N$ and two-particle component $V_N$ and one can retain elements in (\ref{xiex}) that 
are correct through the second order, i.e.,
\begin{equation}
	\Gamma \simeq (H)_{\rm act}+([H_N,\sigma_{\rm ext}])_{\rm act} +\frac{1}{2!}([[F_N,\sigma_{\rm ext}],\sigma_{\rm ext}])_{\rm act} \;.
\label{xiapp}
\end{equation}
An important  aspect related to the  approximate form of the $\Gamma$ 
operator is related to the excitation order included in the $\sigma_{\rm ext}$ operator, 
which can include single, double, triple, etc. many-body components. Since $\sigma_{\rm 
ext}$ is mainly responsible for dynamic correlation effects various iterative and non-
iterative approximations can be used to evaluate these terms. In this paper, we will 
consider an approximation where 
$\sigma_{\rm ext}$  is represented by external components of singly and doubly excited cluster 
operators ($T_{{\rm ext},1}$ and $T_{{\rm ext},2}$), i.e.,
\begin{align}\label{xiappt2}
	\sigma_{\rm ext} &= T_{\rm ext}-T_{\rm ext}^{\dagger}
	\\\nonumber
	&\simeq T_{{\rm ext},1}^{\rm CCSD}+T_{{\rm ext},2}^{\rm CCSD} - T_{{\rm ext},1}^{{\rm CCSD} \dagger}- T_{\rm ext,2}^{{\rm CCSD} \dagger} \;.
\end{align}
For the simplest approximation of $T_{{\rm ext},1}$ and $T_{{\rm ext},2}$ operators one can employ external parts of the 
CCSD $T_1$ and $T_2$ operators.
More sophisticated approximations may involve external cluster amplitudes corresponding 
to single, double, triple, quadruple, etc. excitations obtained from genuine UCC models.  

Since we are interested  in making DUCC formalism amenable for classical/quantum computing, in the 
remaining part of the paper we will derive the algebraic form for the matrix elements defining   $\Gamma$ operator in the
second-quantized form in the  particle-hole and physical vacuum representations. 
The interest in latter representation is 
mainly caused by the fact that most of  the classical (DMRG)/quantum diagonalizers utilize  physical vacuum representation of the second-quantized 
forms of electronic Hamiltonian. In our analysis we will focus our attention on  one- and two-body interactions. 
To derive these formulas we use a combined approach:\\
\subsection{Determination of the algebraic form of $\Gamma$  Eq.(\ref{xiapp}) using particle-hole formalism.}
Applying the particle-hole variant of Wick's theorem to the operator expressions (defining expansion (\ref{xiapp}) for $T_{\rm ext}$ and $T_{\rm ext}^{\dagger}$ 
given by Eq.(\ref{xiappt2}))
\begin{widetext}
\begin{eqnarray}
&& (H_NT_{\rm ext})_{C,{\rm open}} + (T_{\rm ext}^{\dagger}H_N)_{C,{\rm open}} + \frac{1}{2} 
\lbrace 
((F_NT_{\rm ext})_{C,{\rm open}}T_{\rm ext})_{C,{\rm open}}  +
(T_{\rm ext}^{\dagger}(F_N T_{\rm ext})_{C,{\rm open}})_{C,{\rm open}})  \nonumber \\
&&
+ 
((T_{\rm ext}^{\dagger} F_N)_{C,{\rm open}}T_{\rm ext})_{C,{\rm open}}  +
(T_{\rm ext}^{\dagger}   (T_{\rm ext}^{\dagger} F_N)_{C,{\rm open}})_{C,{\rm open}}
\rbrace
\label{scdc}
\end{eqnarray}
\end{widetext}
and retaining terms only through two-body interactions one obtains
\begin{align}\nonumber
\Gamma=&(\Gamma)_{\rm scalar} + \sum_{P,Q} \gamma^{P}_{Q} N[a_P^{\dagger} a_Q]^{\rm ph} 
\\
&\quad +  \frac{1}{4} \sum_{P,Q,R,S}
\gamma^{PQ}_{RS} N[a_P^{\dagger} a_Q^{\dagger} a_S a_R]^{\rm ph} \;,
\label{normal1}
\end{align}
where again  $P,Q,R,S$ designate 
general active spin-orbital indices (in the forthcoming analysis we will designate occupied and virtual active orbitals by $I,J,K,\ldots$ and
$A,B,C,\ldots$, respectively), and $(...)_{C,{\rm open}}$ designates connected and open (i.e., having external lines in diagrammatic representation) 
part of a given operator expression. 
In Eq.(\ref{normal1}) $(\Gamma)_{\rm scalar}$ corresponds to the scalar part of $\Gamma$ in particle-hole representation, i.e.
$\Gamma_{\rm scalar} = \langle\Phi|\Gamma|\Phi\rangle$.
One should also notice that  first and second, third and sixth, and fourth and fifth operators in expression (\ref{scdc}) are pairs of Hermitian conjugate 
operators (for example, $(H_NT_{\rm ext})_{C,{\rm open}}=(T_{\rm ext}^{\dagger}H_N)_{C,{\rm open}}^{\dagger}$).
The utilization of the particle-hole formalism 
helps in keeping track and including a broader class of correlation effects compared to the physical vacuum representation.
The explicit expressions defining $\gamma^{P}_{Q}$ and $\gamma^{PQ}_{RS}$ amplitudes 
are given in Tables \ref{table_expressions1}, \ref{table_expressions2}, and \ref{table_expressions3}. For 
these tables, we assume that a restricted or unrestricted Hartree-Fock (RHF/UHF) reference is employed
(i.e. all non-diagonal elements of the Fock matrix are zero), and as a result the third and
sixth terms in Eq. \ref{scdc} vanish.
\begin{center}
\begin{table*}[t]
\centering
\caption{The algebraic form of the $\gamma^P_Q$ and $\gamma^{PQ}_{RS}$ amplitudes in Eq.(\ref{normal1})
stemming from the $H_N$ term.
}
\begin{tabular}{rlrlrlrl} \hline  \\
\multicolumn{1}{c}{Amplitude}	& \multicolumn{1}{c}{Expression}  &  	
\multicolumn{1}{c}{Amplitude}	& \multicolumn{1}{c}{Expression}  & 
\multicolumn{1}{c}{Amplitude}	& \multicolumn{1}{c}{Expression}  &
\multicolumn{1}{c}{Amplitude}	& \multicolumn{1}{c}{Expression} \\[0.2cm]
\hline\hline \\
\multicolumn{8}{c}{$H_N$ term} \\[0.2cm]
$\gamma^B_A = $ & $f^B_A$ & $\gamma^J_I =$ & $f^J_I$ & 
$\gamma^I_A = $ & $f^I_A$  & $\gamma^A_I =$ & $f^A_I$ \\[0.2cm]
$\gamma^{BC}_{IA} =$ & $v^{BC}_{IA}$ &
$\gamma^{KA}_{IJ} =$  & $v^{KA}_{IJ} $ &
$\gamma^{CI}_{AB} =$  & $v^{CI}_{AB} $ & 
$\gamma^{IJ}_{KA} =$  & $v^{IJ}_{KA} $ \\[0.2cm]
$\gamma^{JB}_{IA} =$ & $v^{JB}_{IA}$ &
$\gamma^{CD}_{AB} =$ & $v^{CD}_{AB}$ &
$\gamma^{KL}_{IJ} =$ & $v^{KL}_{IJ}$ &
$\gamma^{IJ}_{AB} =$ & $v^{IJ}_{AB}$ \\[0.2cm]
$\gamma^{AB}_{IJ} =$ & $v^{AB}_{IJ}$ &
& & &  \\[0.2cm]
\hline\end{tabular}
\label{table_expressions1}
\end{table*}
\end{center}

\begin{center}
\begin{table*}
\centering
\caption{Algebraic form of $\gamma^P_Q$ and $\gamma^{PQ}_{RS}$ amplitudes in Eq.(\ref{normal1}) (continued).
Amplitudes defining $T_{{\rm ext},1}$ and $T_{{\rm ext},2}$ operators are denoted by $s^a_i$ and $s^{ab}_{ij}$
($i,j,\ldots$ and  $a,b,\ldots$ are generic occupied and virtual spin-orbitals, respectively) while amplitudes defining 
$T_{{\rm ext},1}^{\dagger}$ and $T_{{\rm ext},2}^{\dagger}$ are denoted as $s_a^i$ and $s_{ab}^{ij}$, respectively).
By definition of the external parts of $T$ and $T^{\dagger}$, all  $s$-amplitudes that carry active spin-orbital indices only 
disappear. These terms pertain to active spaces that contain all correlated occupied orbitals and a subset of the virtual ones.  For simplicity we assume a restricted/unrestricted Hartree-Fock (RHF/UHF)  reference $|\Phi\rangle$, 
where all non-diagonal Fock matrix elements disappear. The Einstein summation convention is invoked. 
}
\begin{tabular}{rl} \hline  \\
\multicolumn{1}{c}{Amplitude}	& \multicolumn{1}{c}{Expression} 	\\[0.2cm]
\hline\hline \\
\multicolumn{2}{c}{$(H_{N}T_{{\rm ext}})_{C,{\rm open}}$ term} \\[0.2cm]
$\gamma^B_A +=$ & $ - f^M_A s^B_M + v^{MB}_{eA} s^e_M -\frac{1}{2} v^{MN}_{eA} s^{eB}_{MN} $ \\[0.2cm]
$\gamma^J_I +=$ & $ f^J_e s^e_I + v^{MJ}_{eI} s^e_M +\frac{1}{2}v^{MJ}_{ef} s^{ef}_{MI}$ \\[0.2cm]
$\gamma^I_A +=$ & $v^{MI}_{eA} s^e_M $ \\[0.2cm]
$\gamma^A_I +=$ & $f^A_e s^e_I -f^M_I s^A_M + v^{MA}_{eI} s^e_M + f^M_e t^{eA}_{MI} 
           -\frac{1}{2} v^{MN}_{eI} s^{eA}_{MN} + \frac{1}{2} v^{MA}_{ef} s^{ef}_{MI} $ \\[0.2cm]
$\gamma^{BC}_{IA} +=$ &    $ v^{BC}_{eA}s^{e}_{I} 
- v^{MB}_{AI} s^{C}_{M} + v^{MC}_{AI} s^{B}_{M} + f^{M}_{A}s^{BC}_{MI}
- v^{MB}_{eA} s^{eC}_{MI}+v^{MC}_{eA} s^{eB}_{MI}
+\frac{1}{2} v^{MN}_{IA}s^{BC}_{MN} $  \\[0.2cm]
$\gamma^{KA}_{IJ} +=$ & $ v^{KA}_{eJ} s^{e}_{I} -v^{KA}_{eI} s^{e}_{J} 
+v^{MK}_{IJ} s^{A}_{M} + f^{K}_{e}  s^{eA}_{IJ} +\frac{1}{2} v^{KA}_{ef} s^{ef}_{IJ}
-v^{MK}_{eJ} s^{eA}_{MI} + v^{MK}_{eI} s^{eA}_{MJ}$  \\[0.2cm]
$\gamma^{CI}_{AB} +=$ & $  - v^{MI}_{AB} s^{C}_{M} $ \\[0.2cm]
$\gamma^{IJ}_{KA} +=$  & $v^{IJ}_{eA}s^{e}_{K} $ \\[0.2cm]
$\gamma^{JB}_{IA} +=$  & $v^{JB}_{eA}s^{e}_{I}+v^{MJ}_{IA}s^{B}_{M}-v^{MJ}_{eA}s^{eB}_{MI} $ \\[0.2cm]
$\gamma^{CD}_{AB} +=$ & $v^{MC}_{AB}s^{D}_{M}-v^{MD}_{AB}s^{C}_{M}
+\frac{1}{2}v^{MN}_{AB} s^{CD}_{MN}   $  \\[0.2cm]
$\gamma^{KL}_{IJ}  +=  $ & $v^{KL}_{eJ} s^{e}_{I} - v^{KL}_{eI} s^{e}_{J} +\frac{1}{2} v^{KL}_{ef} s^{ef}_{IJ} $ \\[0.2cm]
$\gamma^{AB}_{IJ} +=$ & $ v^{AB}_{eJ} s^{e}_{I} - v^{AB}_{eI} s^{e}_{J} + v^{MA}_{IJ} s^{B}_{M} 
-v^{MB}_{IJ} s^{A}_{M} + f^{A}_{e} s^{eB}_{IJ} - f^{B}_{e} s^{eA}_{IJ}  + f^{M}_{J} s^{AB}_{MI} -f^{M}_{I} s^{AB}_{MJ}$ \\[0.1cm]
             & $+\frac{1}{2} v^{AB}_{ef} s^{ef}_{IJ} + \frac{1}{2} v^{MN}_{IJ} s^{AB}_{MN}-v^{MA}_{eJ} s^{eB}_{MI} 
             + v^{MA}_{eI} s^{eB}_{MJ}+v^{MB}_{eJ} s^{eA}_{MI} - v^{MB}_{eI} s^{eA}_{MJ}$ \\[0.2cm]
\hline\hline \\
\multicolumn{2}{c}{$(T_{{\rm ext}}^{\dagger}H_{N})_{C,{\rm open}}$ term} \\[0.2cm]
$\gamma^A_B +=$ & $ - f^A_M s^M_B + v^{eA}_{MB} s^M_e-\frac{1}{2} v^{eA}_{MN} s^{MN}_{eB}$ \\[0.2cm]  
$\gamma^I_J +=$ & $ f^e_J s^I_e+ v^{eI}_{MJ} s^e_M + \frac{1}{2} v^{ef}_{MJ} s^{MI}_{ef}    $ \\[0.2cm]
$\gamma^A_I +=$ & $v^{eA}_{MI} s^e_M $ \\[0.2cm]
$\gamma^I_A +=$ & $f^e_A s^I_e - f^I_M s^M_A + v^{eI}_{MA} s^e_M+ f^e_M s^{MI}_{eA}
-\frac{1}{2} v^{eI}_{MN} s^{MN}_{eA} + \frac{1}{2} v^{ef}_{MA} s^{MI}_{ef}$  \\[0.2cm]         
$\gamma^{IA}_{BC} +=$ & $ v^{eA}_{BC} s^I_e - v^{AI}_{MB} s^M_C + v^{AI}_{MC} s^M_B + f^A_M s^{MI}_{BC}
-v^{eA}_{MB} s^{MI}_{eC} + v^{eA}_{MC} s^{MI}_{eB} + \frac{1}{2} v^{IA}_{MN} s^{MN}_{BC} $ \\[0.2cm] 
$\gamma^{IJ}_{KA} +=$ & $ v_{KA}^{eJ} s_{e}^{I} -v_{KA}^{eI} s_{e}^{J} 
+v_{MK}^{IJ} s_{A}^{M} + f_{K}^{e}  s_{eA}^{IJ} +\frac{1}{2} v_{KA}^{ef} s_{ef}^{IJ}
-v_{MK}^{eJ} s_{eA}^{MI} + v_{MK}^{eI} s_{eA}^{MJ}$ \\[0.2cm]
$\gamma_{CI}^{AB} +=$ & $  - v_{MI}^{AB} s_{C}^{M} $ \\[0.2cm]
$\gamma_{IJ}^{KA} +=$  & $v_{IJ}^{eA}s_{e}^{K} $ \\[0.2cm]
$\gamma_{JB}^{IA} +=$  & $v_{JB}^{eA}s_{e}^{I}+v_{MJ}^{IA}s_{B}^{M}-v_{MJ}^{eA}s_{eB}^{MI} $ \\[0.2cm]
$\gamma_{CD}^{AB} +=$ & $v_{MC}^{AB}s_{D}^{M}-v_{MD}^{AB}s_{C}^{M}
+\frac{1}{2}v_{MN}^{AB} s_{CD}^{MN}   $  \\[0.2cm]
$\gamma_{KL}^{IJ}  +=  $ & $v_{KL}^{eJ} s_{e}^{I} - v_{KL}^{eI} s_{e}^{J} +\frac{1}{2} v_{KL}^{ef} s_{ef}^{IJ} $ \\[0.2cm]
$\gamma_{AB}^{IJ} +=$ & $ v_{AB}^{eJ} s_{e}^{I} - v_{AB}^{eI} s_{e}^{J} + v_{MA}^{IJ} s_{B}^{M}  
-v_{MB}^{IJ} s_{A}^{M} + f_{A}^{e} s_{eB}^{IJ} - f_{B}^{e} s_{eA}^{IJ}  + f_{M}^{J} s_{AB}^{MI} -f_{M}^{I} s_{AB}^{MJ}$ \\[0.1cm] 
             & $+\frac{1}{2} v_{AB}^{ef} s_{ef}^{IJ} + \frac{1}{2} v_{MN}^{IJ} s_{AB}^{MN}-v_{MA}^{eJ} s_{eB}^{MI}  
             + v_{MA}^{eI} s_{eB}^{MJ}+v_{MB}^{eJ} s_{eA}^{MI} - v_{MB}^{eI} s_{eA}^{MJ}$ \\[0.2cm]
\hline\end{tabular}
\label{table_expressions2}
\end{table*}
\end{center}

\begin{center}
\begin{table*}
\centering
\caption{Algebraic form of $\gamma^P_Q$ and $\gamma^{PQ}_{RS}$ amplitudes in Eq.(\ref{normal1}) (continued).
Amplitudes defining $T_{{\rm ext},1}$ and $T_{{\rm ext},2}$ operators are denoted by $s^a_i$ and $s^{ab}_{ij}$
($i,j,\ldots$ and  $a,b,\ldots$ are generic occupied and virtual spin-orbitals, respectively) while amplitudes defining 
$T_{{\rm ext},1}^{\dagger}$ and $T_{{\rm ext},2}^{\dagger}$ are denoted as $s_a^i$ and $s_{ab}^{ij}$, respectively).
By definition of the external parts of $T$ and $T^{\dagger}$, all  $s$-amplitudes that carry active spin-orbital indices only 
disappear. These terms pertain to active spaces that contain all correlated occupied orbitals and a subset of the virtual ones.  For simplicity we assume a restricted/unrestricted Hartree-Fock (RHF/UHF)  reference $|\Phi\rangle$, 
where all non-diagonal Fock matrix elements disappear. The Einstein summation convention is invoked. 
}
\begin{tabular}{rl} \hline  \\
\multicolumn{1}{c}{Amplitude}	& \multicolumn{1}{c}{Expression}  \\[0.2cm]
\hline\hline \\
\multicolumn{2}{c}{ $\frac{1}{2}(T_{\rm ext}^{\dagger}(F_N T_{\rm ext})_{C,{\rm open}})_{C,{\rm open}}$ term}\\[0.2cm]  
$\gamma^J_I +=$ & $  \frac{1}{2} s^{J}_{e} f^{e}_{f} s^{f}_{I} - \frac{1}{2} s^M_e f^J_M s^e_I 
             - \frac{1}{4} s^{MJ}_{ef} f^{N}_{I} s^{ef}_{MN} +  \frac{1}{2} s^{MJ}_{eg} f^{e}_{f} s^{fg}_{MI}
             + \frac{1}{4} s^{IM}_{ef} f^{N}_{M} s^{ef}_{NJ} $ \\[0.2cm]
$\gamma^A_B +=$ & $ \frac{1}{2} s^{M}_{B} f^{N}_{M} s^{A}_{N} - \frac{1}{2} s^{M}_e f^e_B s^A_M 
             + \frac{1}{4} s^{MN}_{fB}  f^{A}_{e} s^{ef}_{MN} + \frac{1}{4} s^{MN}_{eB} f^{e}_{f} s^{Af}_{MN}
             - \frac{1}{2} s^{MK}_{eB} f^{N}_{K} s^{eA}_{NM}$ \\[0.2cm]
$\gamma^A_I +=$ & $ \frac{1}{2} s^{M}_{e} f^{A}_{f} s^{ef}_{MI} - \frac{1}{2} s^{M}_{e} f^{N}_{I} s^{eA}_{MN}
             - \frac{1}{2} s^{M}_{e} f^{N}_{M} s^{Ae}_{IN} + \frac{1}{2} s^{M}_{e} f^{e}_{f} s^{Af}_{IM} $ \\[0.2cm]
$\gamma^{IJ}_{KL} +=$ & $ \frac{1}{4} s^{IJ}_{ef} f^{M}_{L} s^{ef}_{MK} - \frac{1}{4} s^{IJ}_{ef} f^{M}_{K} s^{ef}_{ML} 
                   + \frac{1}{2} s^{IJ}_{eg} f^{e}_{f} s^{fg}_{KL} $   \\[0.2cm]
$\gamma^{IA}_{JB} +=$ & $\frac{1}{2} s^{MI}_{eB} f^{N}_{J} s^{eA}_{MN} - \frac{1}{2} s^{MI}_{eB} f^{A}_{f} s^{ef}_{MJ}
                   - \frac{1}{2} s^{MI}_{eB} f^{e}_{f} s^{fA}_{MJ} + \frac{1}{2} s^{MI}_{eB} f^{N}_{M} s^{eA}_{NJ} $ \\[0.2cm]
$\gamma^{AB}_{CD} +=$ & $\frac{1}{4} s^{MN}_{CD} f^{A}_{e} s^{eB}_{MN} - \frac{1}{4} s^{MN}_{CD} f^{B}_{e} s^{eA}_{MN}
                   - \frac{1}{2} s^{MK}_{CD} f^{N}_{M} s^{AB}_{NK} $ \\[0.2cm]
$\gamma^{IJ}_{KA} +=$ & $ - \frac{1}{2} s^{IJ}_{eA} f^{M}_{K} s^{e}_{M} $ \\[0.2cm]
$\gamma^{CI}_{AB} +=$ & $ - \frac{1}{2} s^{MI}_{AB} f^{C}_{e} s^{e}_{M} $ \\[0.2cm]
$\gamma^{KA}_{IJ} +=$ & $ \frac{1}{2} s^{K}_{e} f^{e}_{f} s^{fA}_{IJ} - \frac{1}{2} s^{K}_{e} f^{M}_{J} s^{eA}_{IM} 
                   + \frac{1}{2} s^{K}_{e} f^{M}_{I} s^{eA}_{JM} + \frac{1}{2} s^{K}_{e} f^{A}_{f} s^{ef}_{IJ}$ \\[0.2cm]
$\gamma^{AB}_{CI} +=$ & $ \frac{1}{2} s^{M}_{C} f^{N}_{M} s^{AB}_{NI} - \frac{1}{2} s^{M}_{C} f^{B}_{e} s^{eA}_{IM} 
                   + \frac{1}{2} s^{M}_{C} f^{A}_{e} s^{eB}_{IM} + \frac{1}{2} s^{M}_{C} f^{N}_{I} s^{AB}_{MN}$ \\[0.2cm]
\hline\hline \\
\multicolumn{2}{c}{ $\frac{1}{2} ( (T_{\rm ext}^{\dagger} F_N)_{C,{\rm open}}T_{\rm ext})_{C,{\rm open}}  $ term} \\[0.2cm]
$\gamma_J^I +=$ & $  \frac{1}{2} s_{J}^{e} f_{e}^{f} s_{f}^{I} - \frac{1}{2} s_M^e f_J^M s_e^I 
             - \frac{1}{4} s_{MJ}^{ef} f_{N}^{I} s_{ef}^{MN} +  \frac{1}{2} s_{MJ}^{eg} f_{e}^{f} s_{fg}^{MI}
             + \frac{1}{4} s_{IM}^{ef} f_{N}^{M} s_{ef}^{NJ} $ \\[0.2cm]
$\gamma_A^B +=$ & $ \frac{1}{2} s_{M}^{B} f_{N}^{M} s_{A}^{N} - \frac{1}{2} s_{M}^e f_e^B s_A^M 
             + \frac{1}{4} s_{MN}^{fB}  f_{A}^{e} s_{ef}^{MN} + \frac{1}{4} s_{MN}^{eB} f_{e}^{f} s_{Af}^{MN}
             - \frac{1}{2} s_{MK}^{eB} f_{N}^{K} s_{eA}^{NM}$ \\[0.2cm]
$\gamma_A^I +=$ & $ \frac{1}{2} s_{M}^{e} f_{A}^{f} s_{ef}^{MI} - \frac{1}{2} s_{M}^{e} f_{N}^{I} s_{eA}^{MN}
             - \frac{1}{2} s_{M}^{e} f_{N}^{M} s_{Ae}^{IN} + \frac{1}{2} s_{M}^{e} f_{e}^{f} s_{Af}^{IM} $ \\[0.2cm]
$\gamma_{IJ}^{KL} +=$ & $ \frac{1}{4} s_{IJ}^{ef} f_{M}^{L} s_{ef}^{MK} - \frac{1}{4} s_{IJ}^{ef} f_{M}^{K} s_{ef}^{ML} 
                   + \frac{1}{2} s_{IJ}^{eg} f_{e}^{f} s_{fg}^{KL} $   \\[0.2cm]
$\gamma_{IA}^{JB} +=$ & $\frac{1}{2} s_{MI}^{eB} f_{N}^{J} s_{eA}^{MN} - \frac{1}{2} s_{MI}^{eB} f_{A}^{f} s_{ef}^{MJ}
                   - \frac{1}{2} s_{MI}^{eB} f_{e}^{f} s_{fA}^{MJ} + \frac{1}{2} s_{MI}^{eB} f_{N}^{M} s_{eA}^{NJ} $ \\[0.2cm]
$\gamma_{AB}^{CD} +=$ & $\frac{1}{4} s_{MN}^{CD} f_{A}^{e} s_{eB}^{MN} - \frac{1}{4} s_{MN}^{CD} f_{B}^{e} s_{eA}^{MN}
                   - \frac{1}{2} s_{MK}^{CD} f_{N}^{M} s_{AB}^{NK} $ \\[0.2cm]
$\gamma_{IJ}^{KA} +=$ & $ - \frac{1}{2} s_{IJ}^{eA} f_{M}^{K} s_{e}^{M} $ \\[0.2cm]
$\gamma_{CI}^{AB} +=$ & $ - \frac{1}{2} s_{MI}^{AB} f_{C}^{e} s_{e}^{M} $ \\[0.2cm]
$\gamma_{KA}^{IJ} +=$ & $ \frac{1}{2} s_{K}^{e} f_{e}^{f} s_{fA}^{IJ} - \frac{1}{2} s_{K}^{e} f_{M}^{J} s_{eA}^{IM} 
                   + \frac{1}{2} s_{K}^{e} f_{M}^{I} s_{eA}^{JM} + \frac{1}{2} s_{K}^{e} f_{A}^{f} s_{ef}^{IJ}$ \\[0.2cm]
$\gamma_{AB}^{CI} +=$ & $ \frac{1}{2} s_{M}^{C} f_{N}^{M} s_{AB}^{NI} - \frac{1}{2} s_{M}^{C} f_{B}^{e} s_{eA}^{IM} 
                   + \frac{1}{2} s_{M}^{C} f_{A}^{e} s_{eB}^{IM} + \frac{1}{2} s_{M}^{C} f_{N}^{I} s_{AB}^{MN}$ \\[0.2cm]
\hline\end{tabular}
\label{table_expressions3}
\end{table*}
\end{center}
\begin{center}
\begin{table*}
\centering
\caption{Translation of  particle-hole normal product forms for typical strings of creation/annihilation operators into the 
physical-vacuum normal product form (all creation operators are to the left with respect to the annihilation operators).
}
\begin{tabular}{rcl} \hline  \\
\multicolumn{1}{l}{normal product form:}	& \multicolumn{1}{c}{ } & 	\multicolumn{1}{l}{normal product form:}  \\
\multicolumn{1}{l}{particle-hole formalism}	& \multicolumn{1}{c}{ } & 	\multicolumn{1}{l}{physical vacuum formalism}  \\[0.2cm]
\hline\hline \\
$N[a_B^{\dagger} a_A]^{\rm ph} $ &=& $a_{B}^{\dagger} a_A$ \\[0.2cm]
$N[a_J^{\dagger} a_I]^{\rm ph} $ &=& $ a_J^{\dagger} a_I - \delta_{IJ} $  \\[0.2cm]
$N[a_I^{\dagger}  a_A]^{\rm ph} $ &=& $a_I^{\dagger} a_A $ \\[0.2cm]
$N[a_A^{\dagger} a_I]^{\rm ph} $ &=& $a_A^{\dagger} a_I $ \\[0.2cm]
$N[a_B^{\dagger} a_C^{\dagger}  a_A a_I ]^{\rm ph}$ &=& $-a_B^{\dagger} a_C^{\dagger} a_I a_A$ \\[0.2cm]
$N[a_K^{\dagger} a_A^{\dagger} a_J a_I]^{\rm ph} $ &=& $-a_A^{\dagger}a_K^{\dagger} a_J a_I - \delta_{IK}a_A^{\dagger} a_J 
+\delta_{JK} a_A^{\dagger} a_I$ \\[0.2cm]
$N[ a_C^{\dagger} a_I^{\dagger} a_B a_A]^{\rm ph}$  &=& $a_C^{\dagger} a_I^{\dagger} a_B a_A $ \\[0.2cm]
$N[a_I^{\dagger} a_J^{\dagger} a_A a_K]^{\rm ph}$ &=& $-a_I^{\dagger} a_J^{\dagger} a_K a_A  - \delta_{IK} a_J^{\dagger} a_A
+ \delta_{JK} a_I^{\dagger} a_A $ \\[0.2cm]
$N[a_J^{\dagger} a_B^{\dagger} a_A a_I ]^{\rm ph}$ &=& $a_B^{\dagger} a_J^{\dagger} a_I a_A - \delta_{IJ} a_B^{\dagger} a_A $ \\[0.2cm]
$N[a_C^{\dagger} a_D^{\dagger} a_B a_A]^{\rm ph}$ &=& $a_C^{\dagger} a_D^{\dagger} a_B a_A$ \\[0.2cm]
$N[a_K^{\dagger} a_L^{\dagger} a_J a_I]^{\rm ph}$ &=& $a_K^{\dagger} a_L^{\dagger} a_J a_I - \delta_{JL} a_K^{\dagger} a_I
+\delta_{IL} a_K^{\dagger} a_J  + \delta_{JK} a_L^{\dagger} a_I - \delta_{IK} a_L^{\dagger} a_J  + \delta_{JL}\delta_{IK} 
- \delta_{JK} \delta_{IL} $  \\[0.2cm]
$N[a_I^{\dagger} a_J^{\dagger} a_B a_A ]^{\rm ph}$  &=& $a_I^{\dagger} a_J^{\dagger} a_B a_A $ \\[0.2cm]
$N[a_A^{\dagger} a_B^{\dagger} a_J a_I ]^{\rm ph}$ &=& $a_A^{\dagger} a_B^{\dagger} a_J a_I $  \\[0.2cm]
\hline\end{tabular}
\label{table_expressions4}
\end{table*}
\end{center}
\subsection{Determination of the physical-vacuum representation of the $\Gamma$ operator.}

In order to find an equivalent characterization of the $\Gamma$ operator given by Eq.(\ref{normal1}) using 
the physical vacuum parametrization we will employ the set of identities from Table
\ref{table_expressions4}
that translate particle-hole normal product forms for typical strings of creation/annihilation operators to physical vacuum expressions 
where all creation operators are placed to the left of the annihilation operators.  Using these identities the physical vacuum 
Hamiltonian $\Gamma$, in one- and two-body interaction approximation takes the form:
\begin{equation}
\Gamma = \sum_{PQ} \chi^P_Q a_P^{\dagger} a_Q + \frac{1}{4} \sum_{P,Q,R,S} \chi^{PQ}_{RS} a_P^{\dagger} a_Q^{\dagger} a_S a_R \;,
\label{gammaph}
\end{equation}
where all $\chi^P_Q$ and $\chi^{PQ}_{RS}$ coefficients are listed in Table \ref{table_expressions5}. 
These matrix elements can be implemented and used as an input for full configuration interaction  type diagonalizers 
 (in this case limited to the diagonalization in the corresponding  active space) including various "full" CC approaches 
 (CC approaches involving all possible excitations within the 
 active space), density matrix renormalization group, and quantum simulators (employing either QPE or VQE).
 \begin{center}
\begin{table*}
\centering
\caption{The algebraic form of $\chi^P_Q$ and $\chi^{PQ}_{RS}$ amplitudes as functions of 
$\gamma^P_Q$ and $\gamma^{PQ}_{RS}$ ones.
}
\begin{tabular}{rlcrl} \hline  \\
\multicolumn{1}{c}{Amplitude}	& \multicolumn{1}{c}{Expression} & \multicolumn{1}{c}{ $\;\;\;\;\;\;$}& 	\multicolumn{1}{c}{Amplitude} & \multicolumn{1}{c}{Expression} \\[0.2cm]
\hline\hline \\  
$\chi^B_A = $ & $\gamma^B_A-\sum_{M} \gamma^{MB}_{MA}$       & &
$\chi^J_I =$    & $\gamma^J_I  -\sum_{M} \gamma^{MJ}_{MI} $         \\[0.3cm]
$\chi^I_A = $  & $\gamma^I_A - \sum_{M} \gamma^{MI}_{MA}$         & &
$\chi^A_I = $  & $\gamma^A_I-\sum_{M} \gamma^{MA}_{MI} $           \\[0.3cm]
$\chi^{BC}_{IA} = $  & $\gamma^{BC}_{IA}$   & & 
$\chi^{AK}_{IJ} = $   &  $\gamma^{AK}_{IJ} $                                    \\[0.3cm]
$\chi^{CI}_{AB} = $ & $\gamma^{CI}_{AB} $   && 
$\chi^{IJ}_{AK} = $  & $\gamma^{IJ}_{AK} $    				       \\[0.3cm]
$\chi^{BJ}_{AI} = $  & $\gamma^{BJ}_{AI} $    && 
$\chi^{CD}_{AB} = $ & $\gamma^{CD}_{AB} $ 				      \\[0.3cm]
$\chi^{KL}_{IJ} = $ & $\gamma^{KL}_{IJ}$       &&
$\chi^{IJ}_{AB} = $ & $\gamma^{IJ}_{AB}$      				     \\[0.3cm]
$\chi^{AB}_{IJ} = $ & $\gamma^{AB}_{IJ}$      &&  \\[0.2cm]
\hline\end{tabular}
\label{table_expressions5}
\end{table*}
\end{center}

\section{Discussion}

The accuracy of  truncations behind the DUCC(2) formalism is contingent upon several factors:
\begin{itemize}
\item[-] the size of the active space (or equivalently the number of active virtual orbitals included in $\mathfrak{h}$,
\item[-] the accuracy of $\sigma_{\rm ext}$ estimates represented by single and double excitations,
\item[-] the role of missing higher-rank many-body effects in the $\sigma_{\rm ext}$ and $\Gamma$ operators.
\end{itemize}
As stated earlier, the size of the active space may effect the accuracy of $\sigma_{\rm ext}$ amplitudes,
which is a consequence of the fact that utilization of larger spaces, whose choice is driven by the value 
of orbital energies, prevent perturbative denominators from being near singular. For situations where
the energy separation between virtual active and inactive  spin orbitals  is sufficiently large one should 
also expect that the role of higher-rank excitations in  $\sigma_{\rm ext}$ is proportionally smaller. 
Otherwise one needs to  include higher many-body components of $\sigma_{\rm ext}$ and $\Gamma$ operator, for 
example, three- and/or four-body components, i.e., $\sigma_{{\rm ext},3}$,  $\sigma_{{\rm ext},4}$, $\ldots$, 
and $\Gamma_{3}$, $\Gamma_4$, $\ldots$. In such cases, one should also expect that standard single reference 
CC formulations 
including $T_3$ and/or  $T_4$ cluster may not be a viable source of the information about exact $\sigma_{{\rm ext},3}$ 
and/or $\sigma_{{\rm ext},4}$ operators. Instead one should resort to using genuine UCC formulations. For example, 
the $T_{{\rm ext},3}$ amplitudes can be extracted from the  UCC(4) model  discussed in Ref.~\cite{unitary1}, where the 
sufficiency conditions for cluster amplitudes
\begin{eqnarray}
0&=&Q_1[F_N T_1+ V_N T_2]_C|\Phi\rangle, \label{ucc4s} 
\\
0&=&Q_2\Big[V_N+F_NT_2+V_N(T_1+T_2+T_3)\nonumber\\
&&\quad+\frac{1}{2} \left(\frac{1}{2} V_NT_2^2 +  T_2^{\dagger}V_NT_2\right)\Big]_C|\Phi\rangle, \label{ucc4d} \\
0&=&Q_3[F_NT_3 +V_NT_2]_C |\Phi\rangle,  \label{ucc4t}
\end{eqnarray}
where $Q_1$, $Q_2$, and $Q_3$ are projection operators onto singly-, doubly-, and triply excited configurations, allow to generate $T_3$ 
amplitudes in on-the-fly manner avoiding in this way typical memory bottlenecks associated with storing the whole set of $T_3$ amplitudes.

The discussed DUCC formalism also offers a possibility of integrating classical and quantum computations, where the 
CC/UCC calculations for $\sigma_{\rm ext}$ and forming $\chi^P_Q$, $\chi^{PQ}_{RS}$, $\ldots$  amplitudes are performed on classical 
computers while the diagonalization step takes advantage of quantum computing resources. 
For this reason it is instructive to discuss the quantum resources as a function of the number of active orbitals ($N_{\rm act}$) 
and total number of spin orbitals ($N_S$) and rank of many-body effects included in the $\Gamma$ operator
expression \ref{gammaph}.

The specific choice of the active-spave or equivalently sub-system embedding algebra $\mathfrak{h}$ defines 
how efficient is  the process of integrating out remaining degrees of freedom (i.e. the parameters/amplitudes 
defining the $T_{\rm ext}$/$\sigma_{\rm ext}$ operator). 
In particular, the proper choice of $\mathfrak{h}$ will impact the accuracy of low-cost (perturbative) estimates of 
$T_{\rm ext}$. Here we will consider two cases: (1) choice of the $\mathfrak{h}$ based on the energy threshold and 
(2) choice of the $\mathfrak{h}$ based on locality criteria (or equivalently sub-system separability  discussed in  previous 
Sections).

In the first case the active space is chosen in analogy to typical applications of multi-reference methods (CASSCF, 
CASPT2~\cite{andersson1990second,andersson1992second},  
NEVPTn~\cite{angeli2001n,angeli2001introduction}, 
MRMBPT~\cite{hirao1992multireference,finley1995applications,finley1996convergence,
chaudhuri2005comparison,nakano1998analytic,rintelman2005multireference,witek2002intruder,witek2003multireference}, 
and DMRG~\cite{sokolov2017time,guo2018perturbative} methods) where active spaces usually contain high- and low-lying occupied and virtual orbitals. In this situation the first order 
contribution  to  $T_{\rm ext}$ can be for example written as 
\begin{equation}
s^{iJ}_{aB} \simeq \frac{v^{iJ}_{aB}}{\epsilon_{i}+\epsilon_{J}-\epsilon_{a}-\epsilon_{B}} \;,
\label{mp2acte}
\end{equation}
where in this specific example $s$-amplitude contains one active occupied (J)  and one active virtual (B) spinorbital indices, 
we will also assume that in the above example $i$ and $a$ represent inactive occupied and virtual spinorbital indices. 
If active-space orbitals are well separated (energetically) from the remaining orbitals, one can expect that the perturbative 
denominators used to define $T_{\rm ext}$ are much larger than those corresponding to excitations within active space and which are 
determined  in the diagonalization procedure. In this case,the use of perturbative techniques should provide reliable  $T_{\rm ext}$ 
estimates. 

\begin{figure}
	\includegraphics[trim={1cm 0 4cm 0},clip,angle=270, width=0.35\textwidth]{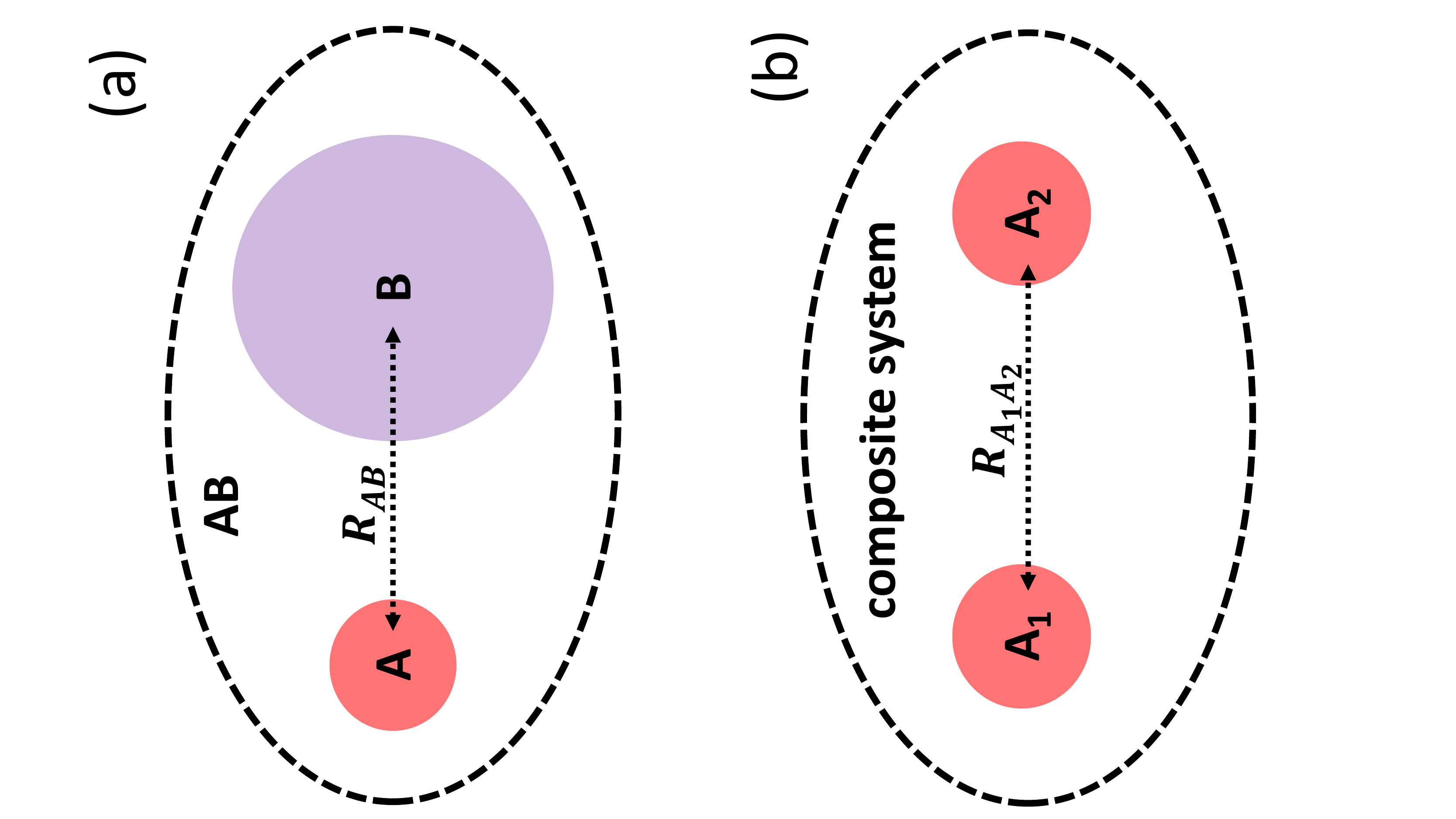}
	\caption{Graphical representation of the active space variant defined by the spatial separation of  two weakly interacting 
	subsystems A and B (a)  and two weakly interacting "active" centers $A_1$ and $A_2$.  Once the active-space orbitals chosen to be localized on sub-system A 
	($A_1$ and $A_2$)  have no significant overlap with remaining  orbitals
	the perturbative arguments apply for numerical estimates of the $T_{\rm ext}$ operator (see text for details).}
\label{fig4}
\end{figure}

In contrast to the energy separation criteria, when  "spatial" arguments are invoked  to define the active space, the 
"smallness" of $\sigma_{\rm ext}$-amplitudes is determined by the decay law of the  orbitals implicated in a specific excitation.
Here, we will consider two specific situations shown in Fig. \ref{fig4}: (a) "active" or strongly correlated sub-system is weakly interacting with 
sub-system B (which can be described by low-order contributions of the many-body perturbation theory) and (b) 
two "active" strongly correlated centers ($A_1$ ad $A_2$) are  embedded in a weakly correlated medium (for example solution). 
In the case (a),  the matrix element in Eq.(\ref{mp2acte})
\begin{equation}
v^{iJ}_{aB} \;,
\label{localv}
\end{equation}
should be "small" since 
spin orbitals $i$ and $a$  vs $J$ and $B$ are spatially well separated (see Fig. \ref{fig4} (a)).
In the second case, one should utilize the joint active spaces  (SES $\mathfrak{h}$)  defined by active orbitals defining sub-systems $A_1$ 
(with corresponding SES $\mathfrak{h}_1$) and $A_2$ (with corresponding SES $\mathfrak{h}_2$), i.e., 
\begin{equation}
\mathfrak{h} = \mathfrak{h}_1 + \mathfrak{h}_2 \;,
\label{hh12}
\end{equation}
where the downfolded Hamiltonian $\bar{H}_{\rm ext}^{\rm DUCC}(\mathfrak{h})$ is given by expression
\begin{equation}
\bar{H}_{\rm ext}^{\rm DUCC}(\mathfrak{h}) = e^{-\sigma_{\rm ext}(\mathfrak{h})} H e^{\sigma_{\rm ext}(\mathfrak{h})} \;.
\label{sumhh}
\end{equation}
Once sub-system $A_1$ and $A_2$ are spatially separated and localized basis set is employed
and $\sigma_{\rm int}(\mathfrak{h}_1)$ and $\sigma_{\rm int}(\mathfrak{h}_2)$ commute
\begin{equation}
[\sigma_{\rm int}(\mathfrak{h}_1),\sigma_{\rm int}(\mathfrak{h}_2)]=0 
\label{commh1h2}
\end{equation}
then further downfolding of 
$\bar{H}_{\rm ext}^{\rm DUCC}(\mathfrak{h})$ is possible, i.e.,
\begin{eqnarray}
\bar{H}_{\rm ext}^{\rm DUCC}(\mathfrak{h}_1) &=& e^{-\sigma_{\rm int}(\mathfrak{h}_2)}
\bar{H}_{\rm ext}^{\rm DUCC}(\mathfrak{h}) e^{\sigma_{\rm int}(\mathfrak{h}_2)}  \label{downs1} \\
\bar{H}_{\rm ext}^{\rm DUCC}(\mathfrak{h}_2) &=& e^{-\sigma_{\rm int}(\mathfrak{h}_1)}
\bar{H}_{\rm ext}^{\rm DUCC}(\mathfrak{h}) e^{\sigma_{\rm int}(\mathfrak{h}_1)}  \label{downs2} \;.
\end{eqnarray}
The above considerations indicate that any  system "separability" parameter can be used to define appropriate model space. 

\section{Resource estimates for simulation on a quantum computer}

The discussed DUCC formalism also offers a possibility of integrating classical and quantum computations, where the 
CC/UCC calculations for $\sigma_{\rm ext}$ and forming $\chi^P_Q$, $\chi^{PQ}_{RS}$, $\ldots$  amplitudes are performed on classical 
computers while the diagonalization step takes advantage of quantum computing resources. 
For this reason it is instructive to discuss the quantum resources as a function of the the number of active orbitals ($N_{\rm act}$) 
and total number of spinorbitals ($N_{\rm S}$) and rank of many-body effects included in the $\Gamma$ operator. 
As an example of the improvement afforded by our downfolding approach, let us evaluate the worst-case resources required by the quantum phase estimation algorithm for obtaining an eigenvalue of the original 
Hamiltonian (involving only one- and two-body interactions) 
 in full space of  all $N_{\rm S}$ orbitals versus DUCC models involving $N_{\rm act}$ active spin orbitals for two situations (2) 
 $\Gamma$ operator contains (1) up to two-body interactions ($\Gamma_1$ and $\Gamma_2$) and (2) up to three-body interactions
 ($\Gamma_1$, $\Gamma_2$, and $\Gamma_3$). 
 
Given an arbitrary unitary $U$ with eigenvalues $U|\psi_j\rangle=e^{i\theta_j}|\psi_j\rangle$ and an input state $|\psi\rangle=\sum_j{\alpha_j}|{\psi_j}\rangle$, quantum phase estimation returns an eigenphase estimate of $\theta_j$ with error $\delta$ for a randomly chosen eigenvector $|{\psi_j}\rangle$ that is sampled with probability $|\alpha_j|^2$. More precisely, the estimate $\hat\theta$ is drawn from the distribution
\begin{align}
    P[\hat\theta=\theta]\propto \left|\sum_{j}\alpha_j f(\theta-\theta_j)\right|^2,
\end{align}
where $f(x)$ is a function that depends on the choice of phase estimation algorithm and is sharply peaked about $x=0$ with width $\delta$. In common variants of phase estimation algorithms~\cite{nielsen2002quantum,Kimmel2015robust,Wiebe2016bayesian}, controlled-$U$ must be applied $\mathcal{O}(1/\delta)$ times to obtain a single $\hat\theta$ estimate, and is the dominant cost.

One common example of $U$ is the real-time evolution operator $e^{-iHt}$, in which case $\theta_j$ is an eigenvalue of $H= \sum^{\mathcal{O}(N_{\rm terms})}_j h_j P_j$, for ${\mathcal{O}(N_{\rm terms})}$ Pauli operators $P_j$ acting on at most $\mathcal{O}(N_{\rm S})$ qubits, and positive coefficients $h_j$, scaled by a constant $t$. Though real-time evolution may be approximated using Trotter-Suzuki product formulas~\cite{berry2007efficient}, it is difficult to obtain tight error bounds on scaling of the approximation error with $N_{\rm act}$. Thus we consider the case where $U$ is a quantum walk~\cite{childs2010relationship,Low2016qubitization,Low2017signal} with eigenvalues commensurate with those of  $e^{\pm i\sin^{-1}(H/\lambda)}$, for $\lambda = \sum_{j} |h_j| \ge \|H\|$.  Unlike Trotter-Suzuki formulas, there is no approximation error in the eigenvalues apart from errors in the Hamiltonian representation. This walk-based approach has become popular of late under the name qubitization~\cite{Low2016qubitization,Low2017signal,babbush2018encoding,Berry2018improved,poulin2018quantum}.  Note that an eigenphase estimate $\hat\theta$ of the quantum walk and its error $\delta$ may be related to the eigenvalue estimate of $H$ by computing $\lambda\sin(\hat\theta)$.  Thus if we wish to learn the eigenvalue within error $\epsilon$ we need to apply the walk operator $\mathcal{O}(\lambda/\epsilon)$ times.

In qubitization, the walk operator is of the form $U={(\rm PREPARE^\dag\otimes I)}\cdot{\rm SELECT}\cdot{(\rm PREPARE\otimes I)}\cdot((1-2|0\cdots0\rangle\langle0\cdots0|)\otimes I)$ for unitary subroutines PREPARE and SELECT. The PREPARE subroutine prepares from the all zero state $|0\cdots0\rangle$ a state of the form $\sum_{j} \sqrt{h_j/\lambda} |j\rangle$ within error $\epsilon$, using $\mathcal{O}(N_{\rm terms})$ quantum gates. Here and in the following, we count the number of arbitrary single- and two-qubit quantum gates. The SELECT subroutine applies each $P_j$ in the Hamiltonian selected by a register that stores the value of $j$. In a naive implementation, this requires $\mathcal{O}(N_{\rm S} N_{\rm terms})$ gates, but optimizations specific to the Jordan-Wigner representation of fermionic Hamiltonians reduce this to $\mathcal{O}(N_{\rm S})$~\cite{babbush2018encoding}, which is subdominant to the cost of the PREPARE circuit.

The overall cost of the algorithm is found by multiplying the cost of the walk operator by the number of iterations needed within phase estimation. Thus the gate complexity of obtaining an estimate of an eigenvalue of $H$ to error $\delta$ is in $\widetilde{\mathcal{O}}(\lambda N_\mathrm{terms}/\delta)$. In the worst-case, we may assume that $N_\mathrm{terms}=\mathcal{O}(N_{\rm S}^4)$, and $\lambda=\mathcal{O}(|H|_\mathrm{max}N_{\rm S}^4)$, where $|H|_\mathrm{max}$ is the maximum absolute value of any entry in the Hamiltonian.

While $|H|_{\mathrm{max}}$ is difficult to estimate in general, there are cases when its scaling can be asymptotically estimated.  If we assume that a local orbital basis is used consisting of atomic orbitals centered at each of the nuclei with charge $Z_i$ then it is straight forward to show that $|h_{pq}|\in \mathcal{O}( \max_i Z_i^2)$ and $|h_{pqrs}|\in \mathcal{O}(\max_i Z_i)$~\cite{babbush2015chemical}.  Therefore, we anticipate that $\lambda \in \mathcal{O}(N_{\rm S}^2 \max_iZ_i^2 + N_{\rm S}^4\max_i Z_i)$ for such problems.  If we consider the nuclear charges to be fixed, we then expect the worst case scaling of a simulation of the total Hamiltonian within error $\delta$ to be in $\mathcal{O}(N_{\rm S}^8/\delta)$ in such cases.

If we consider simulating the downfolded Hamiltonian, one simply replaces $N_\mathrm{terms}=\mathcal{O}(N^4_{\rm act})$ with the number of terms in the downfolded effective Hamiltonian, and similarly for the normalization constant $\lambda=\mathcal{O}(|\Gamma|_\mathrm{max}N^4_{\rm act})$. Importantly, the number of active space orbitals is much smaller than that of the full Hamiltonian, that is $N_{\rm act}\ll N_{\rm S}$. In common practice one normally chooses active spaces such that that amplitude corrections of downfolding are small. In other words, $|\Gamma|_\mathrm{max}$ is expected to be similar to that of the maximum absolute value of $H$ restricted to the active space. If core electrons are also moved out of the active space, one may expect $|\Gamma|_\mathrm{max}\ll |H|_\mathrm{max}$. The case of simulating three-body interactions is less-studied, but the essential idea is identical. With three-body terms, the worst-case $N_\mathrm{terms}=\lambda=\mathcal{O}(N_{\rm act}^6)$, though $\lambda$ could be much smaller if $|\Gamma_3|_\mathrm{max}\ll |\Gamma|_\mathrm{max}$, which would be expected of a small correction.

In practice, it will be essential to understand costs using realistic examples rather than building intuition with the worst-case analysis. In particular, realistic cases might be more accurately captured with low-rank approximations that reduce $N_\mathrm{terms}$ to as small as $\widetilde{\mathcal{O}}(N_{\rm act}^2)$~\cite{motta2018low} for the two-body case -- low-rank approximations for the three-body case are still not well understood. Moreover, various quantum circuit optimization techniques~\cite{Low2018trading} enable a further $\mathcal{O}(N_{\rm act})$ reduction in non-Clifford gate complexity, which are the dominant expensive gates in fault-tolerant quantum computation. More advanced quantum simulation algorithms for structured Hamiltonians could also be applicable, such as those that negate the cost of simulating diagonal terms or exploit large separations in energy scales~\cite{low2018intpicsim, low2018spectral}.

\section{Conclusions}

We have shown that the SES-CC methodology can be extended to unitary CC formalisms which
provides a procedure for downfolding many-body effects into a Hermitian effective Hamiltonian,
in contrast to the earlier coninical SES-CC work in which the effective Hamiltonian was not Hermitian. We
introduced the DUCC model which decouples the classes of excitations used to define the effective
Hamiltonian from those obtained in the corresponding eigenvalue problem. These techniques may 
provide a convenient way of decoupling two types of degrees of freedom corresponding to parameters 
defining low- and high-energy components defining the electronic wave function of interest. 
Using computational chemistry nomenclature these two subsets can be identified with 
static and dynamical correlation effects. However, one can also envision slightly different 
scenarios for the DUCC formalism application where different types of effects (scales) - for 
example,   short- vs  long-range correlations effects - are decoupled  
 using an appropriate form of the local DUCC Ansatz or equivalently adequate definition of the 
 corresponding active space. This development, which provides a rigorous scheme for obtaining Hermitian 
 effective Hamiltonians, opens doors for obtaining the ground-state energy with previously unobtainable 
diagonalization techniques, such as those used in quantum algorithms or the DMRG method. 


Integrating-out the corresponding  fermionic degrees of freedom, which leads to low-dimensionality second-quantized effective Hamiltonians, will open up the possibility of performing quantum simulations on existing quantum simulators, as well as on larger molecular systems. These problems 
will be tested in the forthcoming papers. 
An interesting development area is also  associated with 
recent advances in compressing the second quantized form of 
the electronic Hamiltonian based on the composite Cholesky-SVD decompositions of one- 
and two-electron integrals~\cite{peng2017highly,motta2018low}. One can envisage a further 
extension of the applicability of Cholesky-SVD decomposition to compress $\chi$-amplitudes defining DUCC
downfolded Hamiltonians. 

From the classical computing viewpoint, DUCC downfolding techniques have broader implications and can be  used in the context of the density renormalized group approach (DMRG)~\cite{white1992density,schollwock2005density,dmrg2,chan2011density} by providing a dressed form of the effective Hamiltonian. This approach may 
complement existing perturbative techniques used in the context of the DMRG theory to account for the dynamical correlation effects~\cite{doi:10.1063/1.3275806,kurashige2011second,guo2016n}.

\section{acknowledgement}

This  work  was  supported  by  the "Embedding Quantum Computing into Many-body Frameworks for Strongly Correlated  Molecular and Materials Systems" project, 
which is funded by the U.S. Department of Energy(DOE), Office of Science, Office of Basic Energy Sciences, the Division of Chemical Sciences, Geosciences, and Biosciences.
A portion of this research was funded by the  Quantum Algorithms, Software, and Architectures (QUASAR) Initiative at Pacific Northwest National Laboratory (PNNL). It was conducted under the Laboratory Directed Research and Development Program at PNNL.

%
%
%
%

\end{document}